\documentclass[aps,prb,amsmath,amssymb,floatfix,twocolumn,amsmath,superscriptaddress,twocolumn,nofootinbib,tighten,letterpaper]{revtex4-2}
\usepackage[colorlinks,linkcolor=blue,citecolor=blue,urlcolor=blue]{hyperref}
\usepackage{multirow}
\usepackage{subfigure}
\usepackage{color}
\usepackage{mathrsfs}
\usepackage{hyperref}
\usepackage[normalem]{ulem}
\usepackage{bm}

\usepackage{amssymb}
\usepackage{amsmath}
\renewcommand\vec[1]{\ensuremath\boldsymbol{#1}}

\usepackage{tabularray}
\definecolor{RowColor}{rgb}{0.88,1,0.9}

\usepackage{amsfonts, relsize, color}
\usepackage{graphics}
\usepackage{graphicx}
\usepackage{subfigure}
\usepackage{color}
\usepackage{comment}

\begin{document}
\title{Quantized thermal and spin transports of dirty planar topological superconductors}

\author{Sanjib Kumar Das}
\affiliation{Department of Physics, Lehigh University, Bethlehem, Pennsylvania, 18015, USA}

\author{Bitan Roy}
\affiliation{Department of Physics, Lehigh University, Bethlehem, Pennsylvania, 18015, USA}

\date{\today}

\begin{abstract}
Nontrivial bulk topological invariants of quantum materials can leave their signatures on charge, thermal and spin transports. In two dimensions, their imprints can be experimentally measured from well-developed multiterminal Hall bar arrangements. Here, we numerically compute the low temperature ($T$) thermal ($\kappa_{xy}$) and zero temperature spin ($\sigma^{sp}_{xy}$) Hall conductivities, and longitudinal thermal conductance ($G^{th}_{xx}$) of various prominent two-dimensional fully gapped topological superconductors, belonging to distinct Altland-Zirnbauer symmetry classes, namely $p+ip$ (class D), $d+id$ (class C) and $p \pm ip$ (class DIII) paired states, in mesoscopic six-terminal Hall bar setups from the scattering matrix formalism using Kwant. In both clean and weak disorder limits, the time-reversal symmetry breaking $p+ip$ and $d+id$ pairings show half-quantized and quantized $\kappa_{xy}$ [in units of $\kappa_0=\pi^2 k^2_B T/(3h)$], respectively, while the latter one in addition accommodates a quantized $\sigma^{sp}_{xy}$ [in units of $\sigma^{sp}_0=\hbar/(8 \pi)$]. By contrast, the time-reversal invariant $p \pm ip$ pairing only displays a quantized $G^{th}_{xx}$ at low $T$ up to a moderate strength of disorder. In the strong disorder regime, all these topological responses ($\kappa_{xy}$, $\sigma^{sp}_{xy}$, and $G^{th}_{xx}$) vanish. Possible material platforms hosting such paired states and manifesting these robust topological thermal and spin responses are discussed.     
\end{abstract}

\maketitle

\section{Introduction}

Classification of quantum materials according to the geometry and topology of the underlying fermionic wave-functions, when combined with three non-spatial symmetries, gives rise to the tenfold periodic table of topological phases of matter~\cite{Hasan2010, Qi2011, Altland1997, kanemele2006, BHZ2006, FuKaneMele2007, Fukane2007, moorebalents2007, Schnyder2008, Kitaev2009, rahulroy2009, Ryu2010, vanderbilt2011, Chiu2016}. The participating non-spatial symmetry operations are (a) the time-reversal symmetry (TRS), (b) the antiunitary particle-hole symmetry (PHS), and (c) the unitary particle-hole or chiral or sublattice symmetry (SLS). Among several fascinating features of this topological periodic table of quantum matters, such as the Bott periodicity~\cite{Kitaev2009} and the dimensional reduction~\cite{SCZhang2008} to name a few, a remarkable one is the following. Out of ten possible Altland-Zirnbauer (AZ) symmetry classes, only five are accompanied by non-trivial bulk topological invariants in every dimension. These mathematical quantities (the bulk topological invariants) can nonetheless leave its fingerprints on experimentally measurable transport quantities, and eminently they are expected to be robust against symmetry preserving weak perturbations, such as random impurities. Furthermore, it turns out that such a classification scheme, although tailored for noninteracting fermionic systems, is equally applicable for strongly coupled phases of matter, such as Kondo insulators~\cite{Dzero2010, RoyKondo2014, DzeroReview2016} and superconductors~\cite{Volovik2009, Read2000, Kitaev2006, SCZhang2011PRB, Fujimoto2013, FuAndo2015, Sato2017}, but in terms of emergent weakly correlated Hartree-Fock quasiparticles. Specifically for superconductors, the band topology is computed for weakly interacting emergent neutral Bogoliubov de Gennes (BdG) quasiparticles inside the paired state.

A one-to-one correspondence between the bulk topological invariant and quantized transport quantity is fascinating especially in two spatial dimensions, where they can be directly measured experimentally in multi-terminal Hall bar arrangements. In $d=2$, five topological AZ symmetry classes are A and AII, corresponding to quantum Hall and quantum spin Hall insulators, respectively, and classes D, C and DIII, each of which represents a superconductor. Their symmetry properties are summarized in Table~\ref{table:symmetryclassification}. Here we exclusively focus on two-dimensional topological superconductors. The prominent examples are the (a) TRS breaking $p+ip$ pairing among spinless or equal-spin fermions (class D), (b) TRS breaking spin-singlet $d+id$ pairing (class C), and (c) TRS preserving triplet $p \pm ip$ pairing (class DIII).

Violation of the charge conservation in a superconducting ground state forbids any meaningful measurement of charge transport quantities therein. Therefore, we have to solely rely on the thermal transport (always well defined due to the energy conservation) and in some cases on the spin transport (when the spin rotational symmetry is maintained). Although, the (half-)quantized thermal and spin transport quantities in some of the aforementioned topological paired states are well-appreciated in the literature from the field-theoretic approach and the Kubo formalism~\cite{Read2000, SCZhang2011PRB, Fujimoto2013, senthilmarston1999}, here we compute these quantities in finite size mesoscopic six-terminal Hall bar arrangements from the scattering matrix formalism using the software package Kwant~\cite{Groth2014}, starting from their square lattice-based tight-binding descriptions. Our key findings are summarized below. 

\begin{table}[t!]
\renewcommand{\arraystretch}{1}
\begin{tblr}{width=0.25\columnwidth, colspec = {@{}|Q[c,m,1.2cm]@{}|[white]Q[c,m,0.9cm]@{}|[white]Q[c,m,0.7cm]@{}|[white]Q[c,m,0.7cm]@{}|[white]Q[c,m,0.7cm]@{}|[white]Q[c,m,4.1cm]|@{}},
cell{3,4}{1-6} = {RowColor},row{1-2}={white},}
\hline
\SetCell[r=2]{c}System & \SetCell[r=2]{c}Class & \SetCell[c=3]{c} Symmetries &&& \SetCell[r=2]{c}Examples\\
\hline
&&\SetCell[r=1]{l}TRS&\SetCell[r=1]{l}PHS&\SetCell[r=1]{c}SLS&\\
\hline
\SetCell[r=2]{c}TIs&  A     & 0   & 0 & 0  & \SetCell[r=1]{c} {Quantum Hall insulator}\\
&AII&-1&0&0&\SetCell[r=1]{c}{$\text{Quantum spin Hall insulator}$}\\
\SetCell[r=3]{c}TSCs&  D     & 0   & +1 & 0  & $p+ip$ superconductor \\
&  C     & 0   & -1 & 0  & $d+id$ superconductor \\
&  DIII     & -1   & +1 & 1  & $p \pm ip$ superconductor \\
\hline
\end{tblr}
\caption{Five topologically nontrivial Altland-Zirnbauer symmetry classes in two spatial dimensions, encompassing topological insulators (TIs) and topological superconductors (TSCs)~\cite{Altland1997, Schnyder2008, Ryu2010}. Symmetry transformations of the corresponding effective single-particle Hamiltonian under the time-reversal symmetry (TRS), particle-hole symmetry (PHS) and sublattice symmetry (SLS). Here, $0$ ($1$) implies the absence (presence) of a specific symmetry, while $\pm$ indicate whether the corresponding symmetry operator squares to $\pm 1$, respectively. In the last column, we show one representative physical system from each Altland-Zirnbauer class. 
}~\label{table:symmetryclassification}
\end{table}

\subsection{Key results}

Two-dimensional $p+ip$ and $d+id$ paired states respectively support half-quantized [Fig.~\ref{fig:pwavetsc}] and quantized [Fig.~\ref{fig:dwavethermaltsc}] thermal Hall conductivity $\kappa_{xy}$ [in units of $\kappa_0=\pi^2 k^2_B T/(3h)$] at low temperature ($T$), intimately tied with the first Chern number of the associated effective single-particle BdG Hamiltonian. Here, $h$ ($k_B$) is the Planck's (Boltzmann) constant and $T$ is the temperature of the system. The $d+id$ paired state in addition features quantized spin Hall conductivity $\sigma^{sp}_{xy}$ [Fig.~\ref{fig:dwavespintsc}] in units of $\sigma^{sp}_0=\hbar/(8\pi)$, where $\hbar=h/(2\pi)$. Finally, we show that the topological invariant of a $p \pm ip$ superconductor can only be revealed from the quantized (in units of $\kappa_0$) longitudinal thermal conductance $G^{th}_{xx}$ [Fig.~\ref{fig:pwavespintsc}], as this paired state has net zero first Chern number.

All these (half)-quantized thermal and spin topological responses of planar topological superconductors are shown to be robust (due to a finite bulk gap) in the presence of weak random charge impurities, the dominant source of elastic scattering in any real material. Only in the strong disorder regime $\kappa_{xy}, \sigma^{sp}_{xy}$ and $G^{th}_{xx} \to 0$, when the disorder strength becomes comparable or larger than the bulk topological gap. See panels (d) and (e) of Figs.~\ref{fig:pwavetsc},~\ref{fig:dwavethermaltsc} and~\ref{fig:pwavespintsc}, and panels (c) and (d) of Fig.~\ref{fig:dwavespintsc}. Although these topological responses are insensitive to the pairing amplitude as long as there exists an underlying Fermi surface, in order to minimize the finite size effect and achieve numerical stability, we set it to be \emph{unity}.

\subsection{Organization}

We now outline the organization principle of the rest of the paper. Section~\ref{sec:ppip} is devoted to the discussion on the TRS breaking $p+ip$ superconductor (class D), and its hallmark half-quantized thermal Hall conductivity in the clean and dirty systems. Finite temperature thermal and zero temperature spin Hall conductivities of a class C spin-singlet $d+id$ paired state (both clean and disordered) are discussed in Sec.~\ref{sec:dpid}. Topological responses of class DIII $p \pm ip$ paired state in terms of quantized longitudinal thermal conductance in both clean and dirty systems are given in Sec.~\ref{sec:ppmip}. Concluding remarks and material pertinence of our study are promoted in Sec.~\ref{sec:conclusions}. Additional technical details of the scattering matrix formalism are presented in the Appendix.


\section{Topological $p+ip$ Superconductor}~\label{sec:ppip} 

We embark on the journey with a two-dimensional class D topological system, namely, the TRS breaking $p+ip$ superconductor~\cite{Read2000}. The effective single-particle BdG Hamiltonian for such a system is described by ${\mathcal H}^{p+ip}_{\rm BdG}(\vec{k})={\boldsymbol \tau} \cdot \vec{d}(\vec{k})$, where~\cite{Qi2006} 
\begin{equation}~\label{eq:ppip}
\vec{d}(\vec{k}) = \bigg( t \sin(k_{x}a), t \sin(k_{y}a), m_0-t_0 \sum_{j=x,y} \cos(k_{j}a) \bigg).
\end{equation}
The two-dimensional Pauli matrices ${\boldsymbol \tau}$ operate on the Nambu indices. Throughout, we set $t=t_0=1$, and the
lattice constant $a=1$. The term proportional to $\tau_3$ results in Fermi surfaces near the $\Gamma$ and ${\rm M}$ points of the Brillouin zone (BZ), respectively for $0<m_0/t_0<2$ and $-2<m_0/t_0<0$. The terms proportional to $\tau_1$ and $\tau_2$ capture topological superconductivity in the system, with the pairing amplitude $t$ (set to be unity).

\begin{figure*}[t!]
\includegraphics[width=0.95\textwidth]{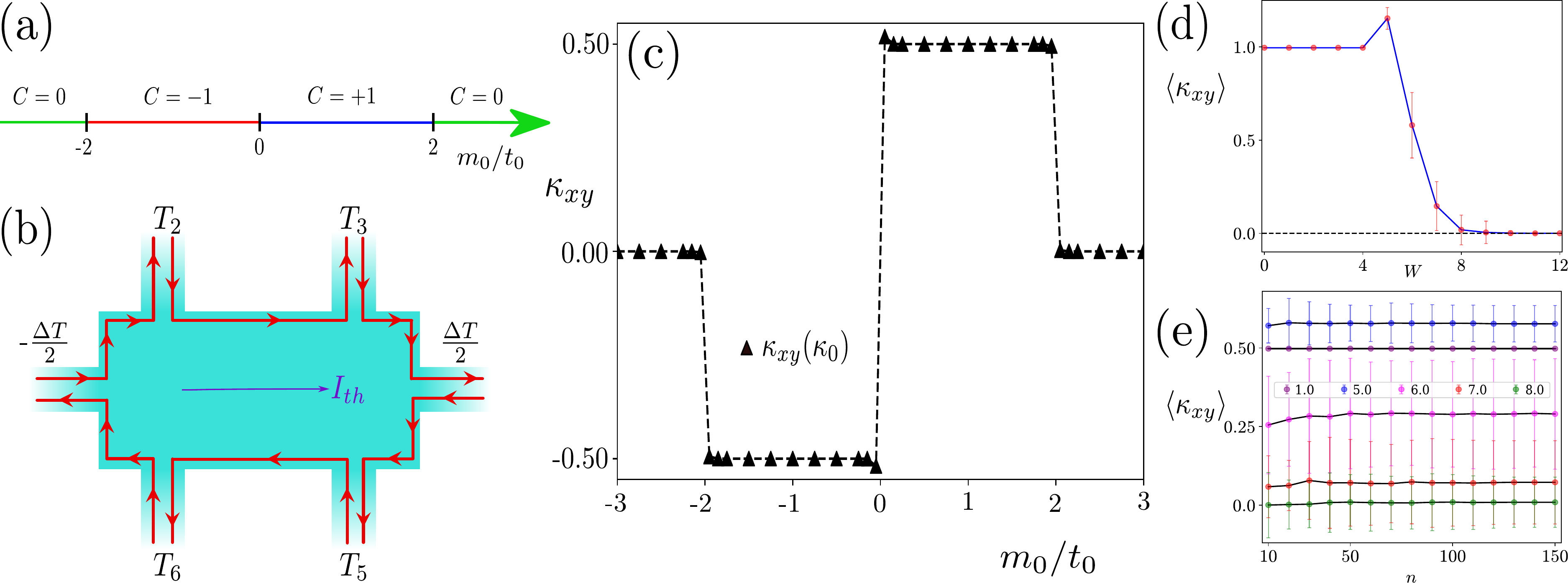}
\caption{(a) The phase diagram for the $p+ip$ superconductor [Eq.~\eqref{eq:ppip}] as a function of $m_0/t_0$ in terms of the first Chern number $C$ [Eq.~\eqref{eq:chernno}]. (b) The six-terminal thermal Hall transport setup for the $p+ip$ superconductor, also showing its chiral Majorana edge mode (red arrow). A longitudinal thermal current ($I_{th}$) flows between lead $1$ and lead $4$. The perpendicular leads serve as the thermal Hall probes, which allow us to calculate the transverse thermal Hall conductivity (THC). (c) The THC ($\kappa_{xy}$), computed for a rectangular system of length $L=200$ and width $D=100$ at the temperature $T=0.01$ as a function of $m_0/t_0$, is half-quantized to $\kappa_{xy}=C/2$ [in the units of $\kappa_0=\pi^{2}k_{B}^2 T/(3h)$] in the topological regime, whereas $\kappa_{xy}=0$ in the trivial regime, where $C=0$. The dotted black lines are guides to the eye. (d) The disorder averaged THC ($\langle \kappa_{xy} \rangle$) as a function of the disorder strength ($W$) for $m_0=t_0=1$ and the pairing amplitude $t=1$, yielding a pairing gap equal to $1$ (in units of $t_0$). Each data point (in red), corresponding to a particular value of $W$, is averaged over 150 independent disorder realizations. (e) The disorder averaged THC as a function of the number of independent disorder realizations ($n$) for various strengths of disorder (mentioned inside the plot) for the same set of parameters as in (d), showing that they saturate after averaging over typically 150 independent disorder realizations. The error bars in (d) and (e) correspond to the standard deviations, which also saturate for $n=150$ (for any $W$).
}~\label{fig:pwavetsc}
\end{figure*}

The superconducting pairing is determined from the electron's spin angular momentum ($\ell$), which forms the Cooper pairs. It can be of spin-singlet (even $\ell$) or spin-triplet (odd $\ell$) in nature. For a system of spinless or spin polarized fermions, the requisite Pauli exclusion principle permits only odd $\ell$ paired states (odd-parity pairing). Here, in Eq.~\eqref{eq:ppip} we consider its simplest example, the TRS breaking $p+ip$ pairing. From the phase diagram of the model, we notice that the topological superconductivity develops in the regime $-2 < m_0/t_0 < 2$ in the presence of an underlying Fermi surface. See Fig.~\ref{fig:pwavetsc}(a). Outside this domain, the system describes a trivial thermal insulator. Within the topological regime, the system supports chiral Majorana edge modes propagating at the boundaries of the system. They manifest in the transport calculation which we will discuss shortly. Since, the formation of the Cooper pairs violates the charge conservation, we rely on the energy conservation principle and compute the thermal Hall conductivity (THC). The THC for this model shows robust half quantization $\kappa_{xy}=\pm 0.5$ when the system falls inside the topological phase, whereas $\kappa_{xy}=0$ otherwise.

We note that the model Hamiltonian for the TRS breaking $p+ip$ superconducting ${\mathcal H}^{p+ip}_{\rm BdG}(\vec{k})$ possesses an anti-unitary PHS, generated by $\Xi=\tau_1 \mathcal{K}$, such that $\{ {\mathcal H}^{p+ip}_{\rm BdG}(\vec{k}), \Xi \}=0$, where $\mathcal{K}$ is the complex conjugation and $\Xi^2=+1$. Note that $\Xi$ exchanges the particle and hole blocks of the Nambu spinor [see Eq.~\eqref{eq:bdgHamilppip}] due to the $\tau_1$ matrix, as well as converts the creation operators to annihilation ones and vice-versa, while taking ${\bm k} \to - {\bm k}$ due to $\mathcal{K}$, as expected from the particle-hole or charge conjugation operator. Additionally, we observe that all three Pauli matrices appear in ${\mathcal H}^{p+ip}_{\rm BdG}(\vec{k})$. Hence, there exists no unitary operator that anticommutes with ${\mathcal H}^{p+ip}_{\rm BdG}(\vec{k})$ (thus no SLS), thereby justifying the class D nature of the system within the AZ classification scheme~\cite{Schnyder2008, Altland1997, Chiu2016, Ryu2010, Kitaev2009}. Therefore, the THC response for this system tracks the value of the first Chern number ($C$), which is defined within the first BZ as
\begin{equation}~\label{eq:chernno}
C=\int_{\rm BZ} \dfrac{d^{2}{\vec k}}{4\pi} \:\: \big[ \partial_{k_x} \hat{\vec{d}}(\vec{k}) \times \partial_{k_y} \hat{\vec{d}}(\vec{k}) \big] \cdot \hat{\vec{d}}(\vec{k}),
\end{equation}
where, $\hat{\vec{d}}(\vec{k})=\vec{d}(\vec{k})/|\vec{d}(\vec{k})|$. Typically, for a TRS breaking topological insulator, the first Chern number is related to the electrical Hall conductivity by the Kubo formula $\sigma_{xy}=C\frac{e^2}{h}$ ($C\in \mathbb{Z}$ group in two dimensions)~\cite{Thouless1982}. Nonetheless, by taking into account the Nambu doubling, which results in a factor of $1/2$ in the BdG Hamiltonian  
\begin{equation}~\label{eq:bdgHamilppip}
H^{p+ip}_{\rm BdG} = \dfrac{1}{2}\sum_{\bm k} \begin{pmatrix}c^{\dagger}_{\bm k} & c_{-{\bm k}} \end{pmatrix} {\mathcal H}^{p+ip}_{\rm BdG} (\vec{k}) \begin{pmatrix} c_{{\bm k}} \\  c^{\dagger}_{-{\bm k}} \end{pmatrix},
\end{equation}
the first Chern number can also be related to the half-quantized THC for a $p+ip$ topological superconductor, which we discuss next. Here, 
$c^{\dagger}_{{\bm k}}$ ($c_{{\bm k}}$) is the fermionic creation (annihilation) operator with momentum ${\bm k}$.

\subsection{Thermal Hall response: Clean $p+ip$ pairing}~\label{sec:pipcleanthermal}

Superconducting system having Cooper pairs does not adhere to the principle of charge conservation, which in turn implies that the electrical Hall conductivity ($\sigma_{xy}$) is a moot quantity. However, we can resort to the energy conservation principle and therefore compute the THC which is a meaningful topological response. We consider a six-terminal Hall bar geometry for the calculation of the THC, as shown in Fig.~\ref{fig:pwavetsc}(b). Six leads are attached to the rectangular scattering region, maintained at a fixed temperature $T$. A longitudinal thermal current ($I_{th}$) then traverses through the system, when a temperature gradient is applied between lead $1$ (at a temperature $T_1=-\Delta T/2$) and lead $4$ (at a temperature $T_4=\Delta T/2$), generating transverse temperatures in the perpendicular leads (namely, lead $2$, lead $3$, lead $5$ and lead $6$), which serve as the temperature leads. With this setup, we note that the current-temperature relation reads as ${\bf I}_{th}={\bf A} {\bf T}$, where ${\bf I}^\top_{th}=(I_{th},0,0,-I_{th},0,0)$, and ${\bf T}^\top=(-\Delta T/2,T_2,T_3,\Delta T/2,T_5,T_6)$. The matrix elements of ${\bf A}$ are calculated from~\cite{Long2011, Fulga2020, Das2023}
\begin{equation}~\label{eq:currtemp}
A_{ij} =  \int_0^\infty \frac{E^2}{T} \left( -\frac{\partial f(E, T)}{\partial E} \right)\left[\delta_{ij} \mu_j-  {\rm Tr}({t}_{ij}^\dagger {t}_{ij}) \right] dE.
\end{equation}
Here, $\mu_j$ represents the number of propagating channels in the $j$th lead, $f(E,T)=1/(1+\exp{[E/(k_{B}T)]})$ denotes the Fermi-Dirac distribution function, $E$ is the energy, $t_{ij}$ is the transmission block of the scattering matrix between the lead $i$ and lead $j$, and the trace (Tr) is performed over all the transmission channels. See the Appendix~\ref{append:scattering} for additional details.

Once we numerically obtain the matrix ${\bf A}$ using Kwant~\cite{Groth2014}, the transverse thermal Hall resistance can be obtained as $R^{th}_{xy}=(T_2+T_3-T_5-T_6)/(2I_{th})$, and the inverse of this quantity is termed the THC, defined as 
\begin{equation}~\label{eq:kappadef}
\kappa_{xy}= \pi^2 k^2_B T/(3h) \left( R^{th}_{xy} \right)^{-1}.
\end{equation}
For all our numerical THC calculations, we set $k_B=h=1$. From the Fig.~\ref{fig:pwavetsc}(c), we notice that the THC remains half-quantized to the values $\kappa_{xy}=\kappa_0 C/2$ in the topological regime with $C= \pm 1$, but vanishes otherwise.

\subsection{Thermal Hall response: Dirty $p+ip$ pairing}~\label{sec:dirtypip}

Topology of insulating systems (electrical and thermal) is robust against weak perturbations that preserve the necessary symmetries of the system or at least when the symmetry is protected on average~\cite{Senthil2000, Nomura2007, Shindou2009, Mirlin2010, Goswami2011, Ringel2012, Kobayashi2013, Fulga2014, ChrisMudry2015, Alavirad2017}. On the other hand, disorder is unavoidably present in real materials, which can in principle be detrimental for the topological features of quantum materials (such as $\kappa_{xy}$). Therefore, we investigate the robustness of the half-quantized $\kappa_{xy}$ of the $p+ip$ paired state against symmetry preserving disorder. We add random charge impurities to each site of the two-dimensional lattice, the dominant source of elastic scattering in any real material, which in the Nambu-doubled basis enter as the on-site \emph{mass} disorder $V(\vec{r}) \tau_3$ for Dirac fermions, featured by the $p+ip$ pairing terms. The quantity $V(\vec{r})$ is uniformly and randomly distributed within the range $[-W/2,W/2]$ for every site belonging to the rectangular scattering region, and $W$ is the disorder strength. We compute the disorder averaged THC $\langle \kappa_{xy} \rangle$ for the class D $p+ip$ superconductor following the prescription mentioned in the previous sections [Sec.~\ref{sec:pipcleanthermal}]. The results are shown in Fig.~\ref{fig:pwavetsc}(d).

The convergence of $\langle \kappa_{xy} \rangle$ is ensured after averaging over a large number of independent disorder configurations $n\sim 150$ (typically), around which $\langle \kappa_{xy} \rangle$ becomes insensitive to $n$ within the numerical accuracy. See Fig.~\ref{fig:pwavetsc}(e). The topological half-quantization of $\langle \kappa_{xy} \rangle$ persists in the weak disorder regime ($W \lesssim 4$), and decays to $\langle \kappa_{xy} \rangle=0$ for large disorder strength ($W \gtrsim 10$), while acquiring non-universal and non-quantized values for intermediate strength of disorder. See Fig.~\ref{fig:pwavetsc}(d). The robustness of the THC in the weak and its disappearance in the strong disorder regimes can be qualitatively understood in the following way. We note that the topological response quantity $\langle \kappa_{xy} \rangle$ is protected by the bulk gap of the BdG fermions. Despite disorder tending to diminish the bulk gap, it continues to protect a half-quantized $\langle \kappa_{xy} \rangle$ for weak disorder. However, in the presence of strong disorder (when $W \gtrsim$ bulk gap), the system becomes a trivial thermal insulator, resulting in $\langle \kappa_{xy} \rangle=0$.

\begin{figure*}[t!]
\includegraphics[width=\linewidth]{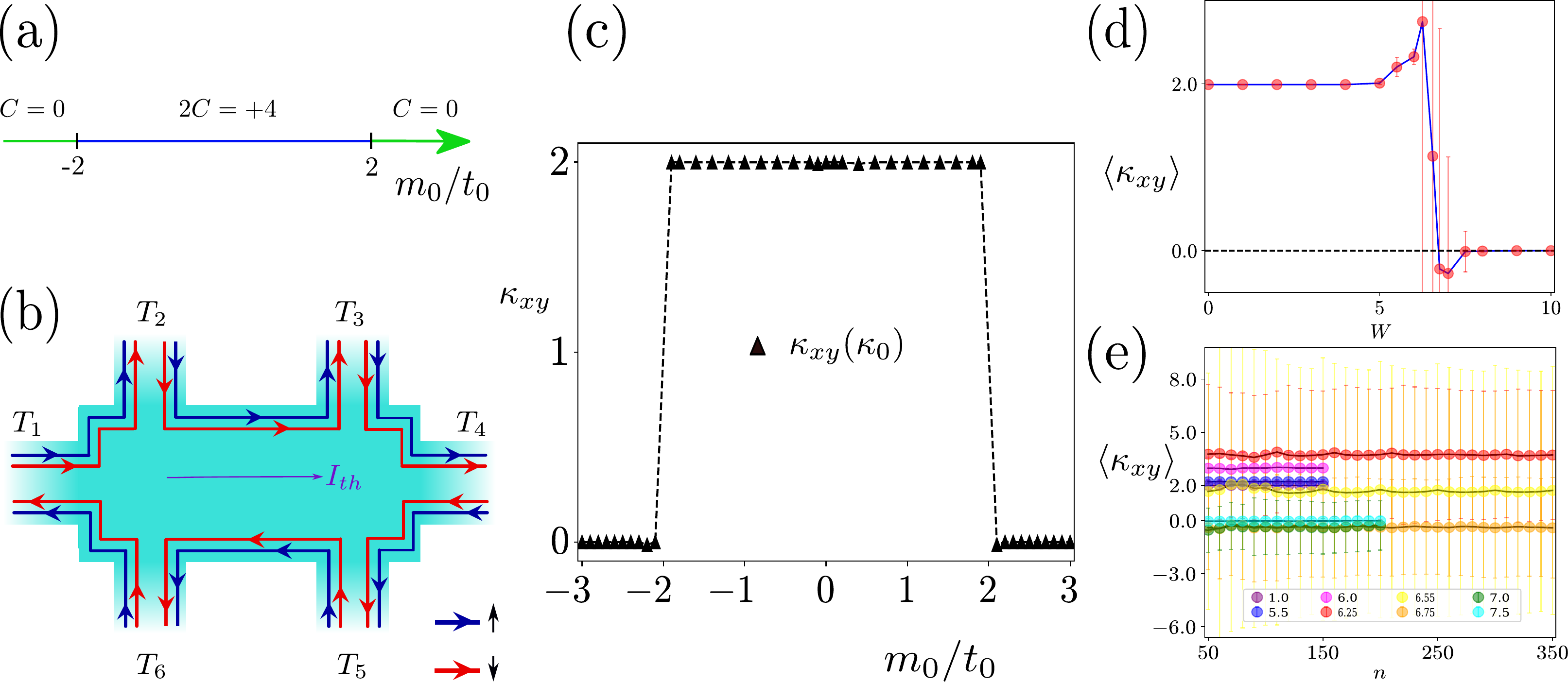}
\caption{(a) The phase diagram for the $d+id$ superconductor [Eq.~\eqref{eq:didbdgmodified}] as a function of $m_0/t_0$ in terms of the first Chern number $2C$ [Eq.~\eqref{eq:chernno}] being equal to $4$ in the topological regime ($|m_0/t_0|<2$) and $0$ otherwise. The factor of $2$ comes from the spin degeneracy of the effective BdG Hamiltonian [Eq.~\eqref{eq:didbdgmodified}]. (b) The six-terminal thermal Hall transport setup for the $d+id$ superconductor, also showing the unidirectional chiral Majorana edge modes for opposite spin projections. A longitudinal thermal current ($I_{th}$) flows between lead $1$ and lead $4$. The perpendicular leads serve as the thermal Hall probes which allow us to calculate the transverse thermal Hall conductivity (THC). (c) The THC ($\kappa_{xy}$) as a function of $m_0/t_0$ is quantized to $2$ (in units of $\kappa_0$) in the topological regime as two unidirectional edge spin channels contribute equally to the THC [see panel (b)], whereas $\kappa_{xy}=0$ in the trivial regime. Here, we compute the THC with a rectangular scattering region of length $L=200$ and width $D=100$ at temperature $T=0.01$ for $t_0=\Delta_0=1$. The dotted black lines are guide to the eye. (d) The disorder averaged THC $\langle \kappa_{xy} \rangle$ as a function of the disorder strength ($W$) for $m_0=t_0=1$ and the pairing amplitude $\Delta_0=1$, yielding a pairing gap equal to $1$ (in units of $t_0$). For each $W$ (red points), $\kappa_{xy}$ is averaged over $150-350$ independent disorder configurations (depending on $W$). (e) The variation of $\langle \kappa_{xy} \rangle$ with the number of independent disorder realizations ($n$) for various strengths of $W$, ensuring its numerical convergence for large $n$ for the same set of parameters as in (d). Error bars in (d) and (e) represent the standard deviations, which also saturate for large $n$ (depends on $W$). 
}~\label{fig:dwavethermaltsc}
\end{figure*}

While computing $\langle \kappa_{xy} \rangle$, we also numerically extract the corresponding standard deviation, shown in Fig.~\ref{fig:pwavetsc}(d) and (e). Notice that for (a) weak disorder when $\langle \kappa_{xy} \rangle=0.5 \kappa_0$ and (b) sufficiently strong disorder when $\langle \kappa_{xy} \rangle=0$, the standard deviations from the mean values are negligible (if at all). By contrast, in the intermediate disorder regime, when $\langle \kappa_{xy} \rangle$ takes non-quantized but finite values, the standard deviation is large. This feature can be appreciated by considering the following specific example. Firstly, notice that for any given disorder realization $\kappa_{xy}=0.5 \kappa_0$ or $\kappa_{xy}=0$. When $\langle \kappa_{xy} \rangle=0.5 \kappa_0$, (almost) all the disorder configurations yield $\kappa_{xy}=0.5 \kappa_0$ and when $\langle \kappa_{xy} \rangle=0$ for (almost) all disorder configurations we find $\kappa_{xy}=0$. As a consequence, the standard deviations in these two regimes are negligible (if they exist at all). However, when, for example, $\langle \kappa_{xy} \rangle \approx 0.25 \kappa_0$ which can be found in the intermediate disorder regime, for half of the disorder configurations we find $\kappa_{xy}=0.5 \kappa_0$, while the rests give $\kappa_{xy}=0$, thereby producing a large standard deviation. However, we ensure that the standard deviations for all reported values of $\langle \kappa_{xy} \rangle$ saturate with respect to the number of disorder configurations ($n$), just like their mean values. This feature is present for the THC [Figs.~\ref{fig:dwavethermaltsc}(d) and \ref{fig:dwavethermaltsc}(e)] and spin Hall conductivity [Fig.~\ref{fig:dwavespintsc}(c) and \ref{fig:dwavespintsc}(d)] for the $d+id$ pairing, and longitudinal thermal conductance for the $p \pm ip$ pairing [Fig.~\ref{fig:pwavespintsc}(d) and \ref{fig:pwavespintsc}(e)]. So, we do not repeat this argument any further.


\section{Topological $d+id$ Superconductor}~\label{sec:dpid} 

Next we investigate the thermal and spin Hall responses of a spin-singlet $d+id$ topological superconductor~\cite{Senthil2000, Read2000, Roy2019}. Notice that, similar to the $p+ip$ superconductors, the $d+id$ pairing also breaks the TRS and lacks the SLS, but retains the anti-unitary PHS ($\Xi$). However, for the $d+id$ paired state $\Xi^2=-1$. The normal state Hamiltonian for spinful fermions, required for their condensation in a spin-singlet channel, is described by
\begin{equation}~\label{eq:did}
\mathcal{H}_{\rm nor}(\vec{k}) = [m_0 - t_0 \big ( \cos(k_{x}a) +\cos(k_{y}a) \big )] \sigma_{0} \equiv d_3(\vec{k}) \sigma_0,
\end{equation}
where the Pauli matrices ${\boldsymbol \sigma}$ encode the spin degrees of freedom. The corresponding two-component spinor reads $\Psi^\top_{\vec{k}}=(c_{\vec{k} \uparrow}, c_{\vec{k} \downarrow})$, where $c_{\vec{k} \sigma}$ is the fermion annihilation operator with momentum $\vec{k}$ and spin projection $\sigma=\uparrow, \downarrow$. Incorporation of the superconductivity into this model amounts to the Nambu doubling of the theory, for which we now define a regular Nambu spinor as $\Psi^\top_{\rm Nam}=(\Psi_{\vec{k}}, \Psi^{*}_{\vec{-k}})$. In this basis, the normal state Hamiltonian takes the following form
\begin{equation}~\label{eq:didnambu}
\mathcal{H}^{\rm Nam}_{\rm nor}(\vec{k}) = d_3(\vec{k}) \tau_{3} \sigma_{0},
\end{equation}
where the newly introduced Pauli matrices ${\boldsymbol \tau}$ operate on the Nambu or particle-hole index.

In such a system, in this section, we focus on the even parity pairings with the pairing amplitude $\Delta(\vec{k})$, such that $\Delta(-\vec{k})=\Delta(\vec{k})$. Two prominent choices which satisfy the even parity criterion are the $s$-wave pairing with $\Delta(\vec{k})=\Delta_{0}$ (constant) and $\ell=0$, and the $d$-wave pairing with $\Delta(\vec{k})=\Delta_{0}[\cos(k_{x}a) - \cos(k_{y}a)] \equiv d_1(\vec{k})$ and $\Delta_0 \sin(k_{x}a)\sin(k_{y}a)\equiv d_2(\vec{k})$, corresponding to $\ell=2$, where $\Delta_0$ is the pairing amplitude. These two terms respectively stand for the lattice regularized version of the $d_{x^2-y^2}$ and $d_{xy}$ pairings. Here, we only consider the $d$-wave pairings, as the fully gapped uniform $s$-wave counterpart is topologically trivial. Note that the requirement of the Pauli exclusion principle demands that the spin-singlet pairing terms must appear with the $\sigma_2$ matrix in the spin space, which by virtue of $\sigma^\top_2=-\sigma_2$ ensures the antisymmetric nature of the pairing wave function. The effective single-particle BdG Hamiltonian for the $d+id$ pairing then takes a compact form 
\begin{equation}~\label{eq:didbdg}
\mathcal{H}^{d+id}_{\rm BdG} (\vec{k}) = d_1(\vec{k}) \tau_{1} \sigma_{2} + d_2(\vec{k}) \tau_{2}\sigma_{2} + d_3(\vec{k}) \tau_{3}\sigma_{0},
\end{equation} 
with the components of the $\vec{d}$-vector already announced in this section. The effective BdG Hamiltonian can be cast in a more elegant form by defining a slightly decorated Nambu-doubled basis $\left( \Psi^{\rm dec}_{\rm Nam}\right)^\top=(\Psi_{\vec{k}}, \sigma_2 \Psi_{-\vec{k}}^*)$ by absorbing the unitary part of the time-reversal operator ($\sigma_2$) in the hole sector of the spinor. In this basis 
\begin{equation}~\label{eq:didbdgmodified}
\mathcal{H}^{d+id}_{\rm BdG} (\vec{k}) = d_1(\vec{k}) \tau_{1} \sigma_{0} + d_2(\vec{k})  \tau_{2}\sigma_{0} + d_3(\vec{k}) \tau_{3}\sigma_{0}.
\end{equation} 
Appearance of the Pauli matrix $\sigma_0$ in the spin sector unfolds the singlet nature of the $d+id$ paired state. Thus, the spin degrees of freedom leads to a mere doubling of the BdG Hamiltonian, endowing an SU(2) spin rotational symmetry to $\mathcal{H}^{d+id}_{\rm BdG}(\vec{k})$, generated by $\tau_0 {\boldsymbol \sigma}$. Notice that $\mathcal{H}^{d+id}_{\rm BdG}(\vec{k})$ enjoys an antiunitary PHS, generated by $\Xi=\tau_2 \sigma_0 {\mathcal K}$, such that $\{ \mathcal{H}^{d+id}_{\rm BdG}(\vec{k}), \Xi \}=0$ and $\Xi^2=-1$. Here as well $\Xi$ exchanges the particle and hole blocks of the Nambu spinor ($\Psi^{\rm dec}_{\rm Nam}$) due to the $\tau_2$ matrix, and converts the creation operators to annihilation ones and vice-versa, while taking ${\bm k} \to - {\bm k}$ due to $\mathcal{K}$ without altering the spin projection (due to $\sigma_0$), as expected from the particle-hole or charge conjugation operator. But, there is no unitary operator that anticommutes with $\mathcal{H}^{d+id}_{\rm BdG}(\vec{k})$, which is thus devoid of the sublattice or chiral symmetry. Hence, the $d+id$ paired state belongs to the AZ class C.

\subsection{Thermal Hall responses of $d+id$ pairing}

Since after a suitable unitary rotation, each term in the effective BdG Hamiltonian $\mathcal{H}^{d+id}_{\rm BdG}$ is accompanied by the identity matrix $\sigma_0$ in the spin sector, the first Chern number ($C$) associated with the $d+id$ paired state can be directly computed from Eq.~\eqref{eq:chernno} in terms of the components of the appropriate $\vec{d}$-vector appearing in the $2 \times 2$ BdG Hamiltonian involving only the ${\boldsymbol \tau}$ matrices (Nambu degrees of freedom), and it is given by $2C$. The extra factor of $2$ arises from the mere doubling of the Hamiltonian in the spin part [see Eq.~\eqref{eq:didbdgmodified}]. Within the entire topological regime, namely $-2<m_0/t_0<2$, the first Chern number associated with $\mathcal{H}^{d+id}_{\rm BdG}$ is thus $2C=4$, while it is trivial ($C=0$) for $|m_0/t_0|>2$. The resulting phase diagram is shown in Fig.~\ref{fig:dwavethermaltsc}(a).

The nontrivial Chern number leaves its signature on the transverse THC, which in this case reads as (following the discussion from Sec.~\ref{sec:ppip})
\begin{equation}~\label{eq:kappadwavethermal}
\kappa_{xy} = C \times \frac{1}{2} \times 2 \times \bigg(\frac{\pi^{2}k_{B}^2 T}{3h} \bigg) \equiv C \kappa_0.
\end{equation} 
The factor of $1/2$ compensates the Nambu doubling, and the factor of $2$ arises due to the spin degeneracy. The THC of a clean $d+id$ paired state can readily be computed in a six-terminal setup, as shown in Fig.~\ref{fig:dwavethermaltsc}(b), employing the same method we previously discussed for the $p+ip$ superconductor in Sec.~\ref{sec:pipcleanthermal}. See Appendix~\ref{append:scattering}. The results are shown in Fig.~\ref{fig:dwavethermaltsc}(c). Indeed, we find $\kappa_{xy}=2$ (in units of $\kappa_0$) in the topological regime, otherwise $\kappa_{xy}=0$.

We also test the robustness of the quantized nature of $\kappa_{xy}$ in the $d+id$ state against a symmetry preserving disorder. It is important to emphasize that the computation of $\kappa_{xy}$ is performed separately on the individual $2\times 2$ blocks involving the ${\boldsymbol \tau}$ matrices only. We add an on-site disorder (random charge scatterer) term $V(\vec{r})\tau_3$, which is drawn from a uniform box distribution in the range $[-W/2,W/2]$ for every site belonging to the real space lattice, and $W$ is the disorder strength. Then we follow the identical steps highlighted in Sec.~\ref{sec:dirtypip} to compute the disorder averaged THC $\langle \kappa_{xy} \rangle$. Once again we find that $\langle \kappa_{xy} \rangle$ retains a quantized value of 2 for small to moderate disorder strength ($W \lesssim 5$) and drops to $\langle \kappa_{xy}\rangle=0$ eventually for large disorder values ($W \gtrsim 8$). See Fig.~\ref{fig:dwavethermaltsc}(d). Since $\langle \kappa_{xy} \rangle$ falls rather rapidly in the intermediate disorder range, we consider a finer mesh in disorder values in this regime. Typically, $\langle \kappa_{xy} \rangle$ becomes insensitive to the number of independent disorder realizations for $n \sim 150-350$ (depending on $W$), as shown in Fig.~\ref{fig:dwavethermaltsc}(e). For each reported value of $\langle \kappa_{xy} \rangle$, we also compute the corresponding standard deviation, shown in  Fig.~\ref{fig:dwavethermaltsc}(d) and (e), which also saturates for the same range of $n$, while displaying qualitatively similar features, previously found and discussed for the $p+ip$ pairing.


\begin{figure*}[t!]
\includegraphics[width=\linewidth]{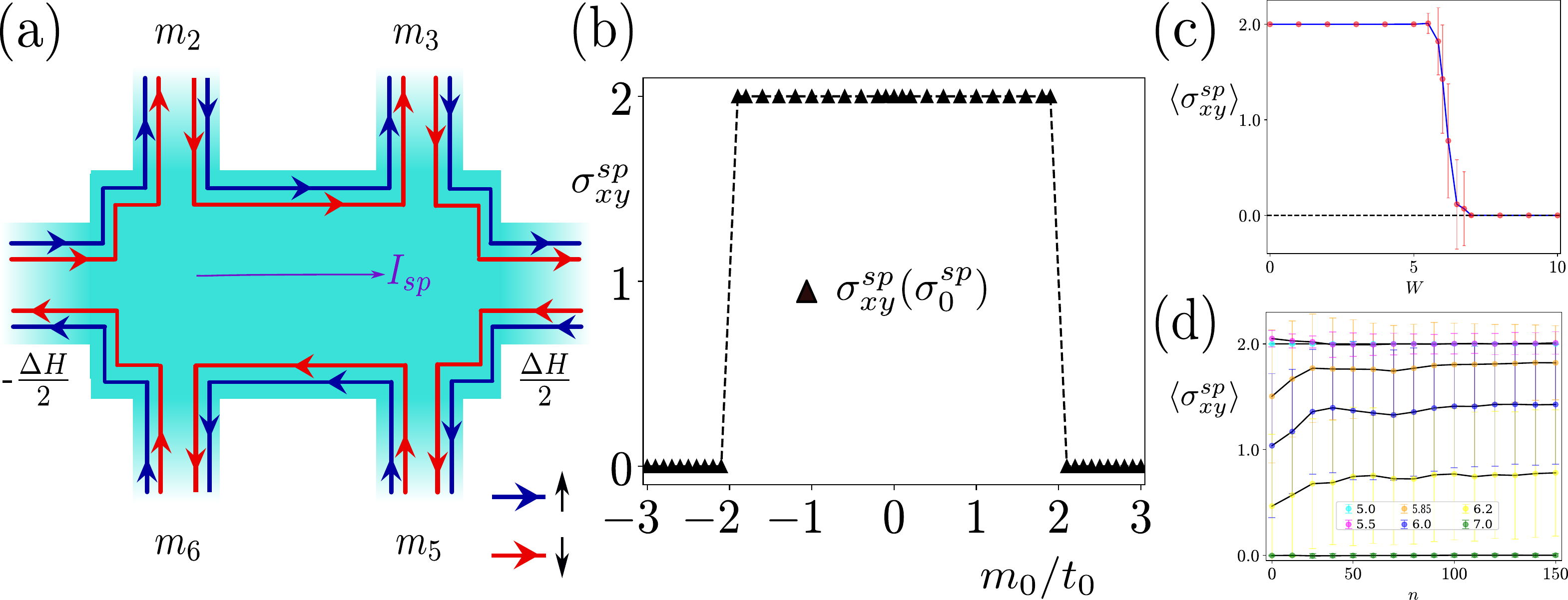}
\caption{(a) The six-terminal spin Hall transport setup for the $d+id$ superconductor at zero temperature, also showing the unidirectional spin degenerate chiral Majorana edge modes. A longitudinal spin current ($I_{sp}$) flows between lead $1$ and lead $4$, due to a magnetic field bias between them. The perpendicular leads work as the spin Hall or the magnetization probes which allow us to calculate the transverse spin Hall conductivity (SHC). (b) The SHC ($\sigma^{sp}_{xy}$) as a function of $m_0/t_0$ is quantized to $2$ [in units of $\sigma^{sp}_0=\hbar/(8\pi)$] in the topological regime, whereas yields $0$ in the trivial regime. Here, we compute $\sigma^{sp}_{xy}$ in a rectangular system of length $L=200$ and width $D=100$, for $t_0=\Delta_0=1$. The dotted black lines are guides to the eye. (c) The disorder averaged SHC $\langle \sigma^{sp}_{xy} \rangle$ as a function of the disorder strength $W$ shows the survival of its robust quantization up to a moderate disorder, while eventually decaying to $\langle \sigma^{sp}_{xy} \rangle=0$ for large $W$. Here the results are obtained for $m_0=t_0=1$ and the pairing amplitude $\Delta_0=1$, yielding a pairing gap equal to $1$ (in units of $t_0$). (d) The same quantity $\langle \sigma^{sp}_{xy}\rangle$ for the same set of parameters as in (c) is plotted with the number of independent disorder realizations ($n$) for a few $W$, showing that it saturates for large $n \sim 150$ ensuring the numerical convergence. The error bars in (c) and (d) stand for the standard deviation, which for any $W$ saturate for $n=150$.
}~\label{fig:dwavespintsc}
\end{figure*}

\subsection{$d+id$ pairing: Spin Hall Conductivity}~\label{subsec:dwavespinhall}

So far the spin degrees of freedom only doubled the amplitude of $\kappa_{xy}$ for the $d+id$ pairing. The broken TRS and the spin-singlet nature of this paired state also manifest quantized spin Hall conductivity, realized at the cost of the spin SU(2) symmetry by applying a \emph{weak} (much smaller than $H_c$ or $H_{c1}$, depending on whether the pairing is type-I or type-II) external magnetic field. Then the external magnetic field ($\vec{H}$) only couples to the spin degrees of freedom (no orbital coupling due to the Meissner effect) via the Zeeman term, which reads as $(\hbar/2) \vec{H} \cdot {\boldsymbol \sigma}$, where $\hbar/2=h/(4\pi)$ plays the role of the magnetic charge (analogous to $e$ being the electrical charge). In the decorated Nambu basis ($\Psi^{\rm dec}_{\rm Nam}$), the Zeeman term reads as $(\hbar/2) \tau_0 \vec{H} \cdot {\boldsymbol \sigma}$. Without any loss of generality, we chose the spin quantization $z$ axis along the external magnetic field, yielding $\vec{H}=(0,0,H)$, and then the Zeeman term takes a simpler form $(\hbar/2) H \tau_0 \sigma_3$. As the zero-energy edge modes of fermionic BdG quasiparticles of the $d+id$ paired state propagate in the same direction for opposite spin projections [see Fig.~\ref{fig:dwavespintsc}(a)], a magnetic field bias between the horizontal leads causes a spin current between them and a magnetization develops between the vertical leads, as we explain below in detail, giving rise to a quantized spin Hall conductance solely due to these edge modes. Therefore, at zero temperature the the spin Hall conductivity of the $d+id$ paired state is given by 
\begin{equation}
\sigma^{sp}_{xy}= \frac{(\hbar/2)^2}{h} \times \frac{1}{2} \times C =\frac{\hbar}{8\pi} C \equiv \sigma^{sp}_0 C, 
\end{equation}    
where the factor of $1/2$ compensates the Nambu doubling, $\sigma^{sp}_0=\hbar/(8\pi)$ is the quantum of the spin Hall conductance, and $C$ is the first Chern number of the $d+id$ paired state, computed from the $2 \times 2$ BdG Hamiltonian involving only the ${\boldsymbol \tau}$ matrices~\cite{senthilmarston1999, Read2000, Roy2019}. Throughout, we compute the spin Hall conductivity in units of $\hbar/(8\pi)$. Therefore, within the topological regime $|m_0/t_0|<2$, we expect $\sigma^{sp}_{xy}=2$, while $\sigma^{sp}_{xy}=0$ for $|m_0/t_0|>2$, which we next confirm from a six-terminal setup.

The six-terminal transport geometry for the computation of the spin Hall conductivity ($\sigma^{sp}_{xy}$) is depicted in Fig.~\ref{fig:dwavespintsc}(a). A magnetic bias field ($-\Delta H/2, \Delta H/2$) is applied between the lead 1 and lead 4, due to which a longitudinal spin current ($I_{sp}$) flows across the system (the scattering region). Such a spin current results in generating different $z$ directional magnetization values ($m_j$s) in the transverse leads. They can be obtained from the spin current-magnetization relation ${\bf I}_{sp}={\bf G}_{sp} {\bf M}$, where ${\bf I}^\top_{sp}=(I_{sp},0,0,-I_{sp},0,0)$ and ${\bf M}^\top=(-\Delta H/2, m_2, m_3, \Delta H/2, m_5, m_6)$. The spin conductance matrix ${\bf G}_{sp}$  contains only the transmission block of the scattering matrix, which we extract using Kwant~\cite{Groth2014}. See Appendix~\ref{append:scattering} for additional details. Subsequently, we compute the transverse spin Hall resistance $R^{sp}_{xy}=(m_2+m_3-m_5-m_6)/(2 I_{sp})$, which leads to the spin Hall conductance $\sigma^{sp}_{xy}=\left( R^{sp}_{xy} \right)^{-1}$. The results are shown in Fig.~\ref{fig:dwavespintsc}(b), confirming $\sigma^{sp}_{xy}=2$ and $0$, respectively, in the topological and trivial regimes. Next we scrutinize the robustness of quantized $\sigma^{sp}_{xy}$ in the presence of random charge impurities.

\begin{figure*}[t!]
\includegraphics[width=\linewidth]{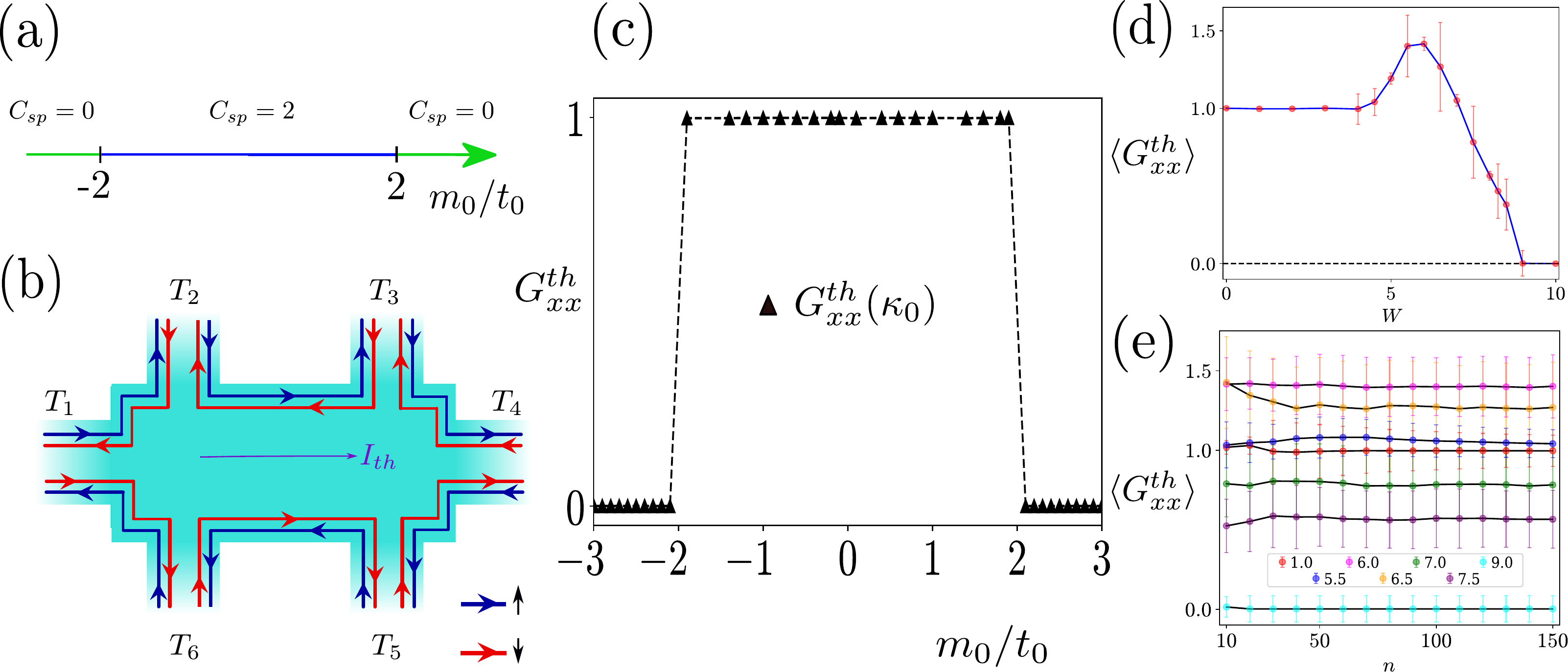}
\caption{(a) The phase diagram of the $p \pm i p$ superconductor [Sec.~\ref{sec:ppmip}] as a function of $m_0/t_0$, showing the topologically trivial and nontrivial regimes in terms of the spin Chern number $C_{sp}$ [Eq.~\eqref{eq:spinchern}]. (b) The six-terminal thermal transport setup for the $p \pm i p$ superconductor, showing the counter-propagating helical Majorana edge modes for opposite spin projections. A longitudinal thermal current ($I_{th}$) flows between lead $1$ and lead $4$. Here, the temperatures in the pair of horizontal probes (namely between lead 2 and lead 3 or lead 5 and lead 6) allow us to calculate the longitudinal thermal conductance ($G^{th}_{xx}$). (c) The $G^{th}_{xx}$ as a function of $m_0/t_0$ is quantized to $1$ (in units of $\kappa_0$) in the topological regime, whereas it yields $0$ in the trivial regime. The computation of $G^{th}_{xx}$ is perform with a rectangular scattering regime of length $L=120$ and width $D=60$, at a temperature $T=0.01$ and for $t=t_0=1$. The dotted lines are guides to the eye. (d) The disorder averaged longitudinal thermal conductance $\langle G^{th}_{xx} \rangle$ with varying disorder strength $W$ for $m_0=t_0=1$ and the pairing amplitude $t=1$, yielding a pairing gap equal to $1$ (in units of $t_0$). For each disorder strength, we average over $n = 150$ independent disorder realizations, for which $\langle G^{th}_{xx} \rangle$ becomes independent of $n$. See the panel (e). The error bars in (d) and (e) indicating the standard deviation saturate for $n=150$ for any disorder strength $W$.
}~\label{fig:pwavespintsc}
\end{figure*}

With the same motivation as in the earlier models, we now investigate the robustness of the quantized $\sigma_{xy}^{sp}$ against a symmetry preserving disorder term, namely the random charge impurities. The computation of the disorder averaged spin Hall conductance $\langle \sigma_{xy}^{sp} \rangle$ now involves on-site disorder term $V(\vec{r})\tau_3$ for each spin projection. The quantity $V(\vec{r})$ is uniformly and independently distributed in the range $[-W/2,W/2]$ for every site belonging to the lattice, and $W$ is the disorder strength. The disorder averaged spin Hall conductance $\langle \sigma_{xy}^{sp} \rangle$ showcases the robustness of this topological response against weak and moderate disorder, which eventually decays to zero in the strong disorder regime. See Fig.~\ref{fig:dwavespintsc}(c). While computing $\langle \sigma_{xy}^{sp} \rangle$, we typically average over $150$ independent disorder realizations for each value of $W$. As the number of disorder realizations ($n$) is increased, the values of $\langle \sigma_{xy}^{sp} \rangle$ get saturated around $n \sim 150$. See Fig.~\ref{fig:dwavespintsc}(d). The standard deviation for each reported value of $\langle \sigma_{xy}^{sp} \rangle$ saturates after averaging over 150 disorder realizations, as shown in Fig.~\ref{fig:dwavespintsc}(c) and (d). Otherwise, its behavior in the weak, moderate and strong disorder regimes are qualitatively similar to the ones, we previously observed for $\langle \kappa_{xy} \rangle$ of $p+ip$ and $d+id$ paired states.


\section{Topological $p \pm ip$ pairing}~\label{sec:ppmip}

Finally, we turn to the situation when a spin degenerate Fermi surface [Eq.~\eqref{eq:did}], discussed in the previous section, becomes susceptible toward the nucleation of a TRS preserving spin-triplet $p \pm i p$ paired state. As we will show shortly that this system besides the TRS, also preserves the anti-unitary particle-hole and the unitary sublattice or chiral symmetry, and belongs to class DIII in the AZ classification scheme. The effective single-particle BdG Hamiltonian for the $p \pm i p$ paired state takes the form ${\mathcal H}^{p \pm i p}_{\rm BdG}(\vec{k})={\boldsymbol \Gamma} \cdot \vec{d}(\vec{k})$ with the $\vec{d}$-vector already given in Eq.~\eqref{eq:ppip}. The mutually anticommuting $4 \times 4$ Hermitian Dirac $\Gamma$ matrices, involving the spin and Nambu indices, take the explicit form $\Gamma_{1}=\sigma_{0}\otimes\tau_{1}, \Gamma_{2}=\sigma_{3}\otimes\tau_{2}$, and $\Gamma_{3}=\sigma_{0}\otimes\tau_{3}$. The Pauli matrices ${\boldsymbol \sigma}$ (${\boldsymbol \tau}$) operate on the spin (Nambu) sector.

The TRS of ${\mathcal H}^{p \pm i p}_{\rm BdG}(\vec{k})$ is generated by ${\mathcal T}=(\sigma_2 \otimes \tau_3) {\mathcal K}$, such that $[{\mathcal T}, {\mathcal H}^{p \pm i p}_{\rm BdG}(\vec{k})]=0$ and ${\mathcal T}^2=-1$, as it should be for spinful fermions. Its anti-unitary PHS is generated by $\Xi= (\sigma_0 \otimes \tau_1) {\mathcal K}$, such that $\{ \Xi, {\mathcal H}^{p \pm i p}_{\rm BdG}(\vec{k}) \}=0$ and $\Xi^2=+1$. Notice that $\Xi$ exchanges the particle and hole blocks of the Nambu spinor due to the $\tau_1$ matrix, and converts the creation operators to annihilation ones and vice-versa, while taking ${\bm k} \to - {\bm k}$ due to $\mathcal{K}$ without affecting the spin projection (due to $\sigma_0$), as expected from the particle-hole or charge conjugation operator. Finally, there are two unitary operators, namely $\Gamma_4=\sigma_1 \otimes \tau_2$ and $\Gamma_5=\sigma_1 \otimes \tau_2$, such that $\{\Gamma_j, {\mathcal H}^{p \pm i p}_{\rm BdG}(\vec{k}) \}=0$ and $\Gamma^2_j=+1$ for $j=4$ and $5$. They generate the SLS of ${\mathcal H}^{p \pm i p}_{\rm BdG}(\vec{k})$. Hence, the $p \pm ip$ paired state belongs to class DIII.

Physically, the model Hamiltonian ${\mathcal H}^{p \pm i p}_{\rm BdG}(\vec{k})$ can be understood as the superposition of the $p+ip$ pairing for the spin up component and the $p-ip$ pairing for the spin down component. As the opposite spin projections are Kramers partners of each other, the TRS of the resulting $p \pm ip$ pairing pins its first Chern number to zero and due to the helical nature of the edge modes, the resultant THC is also zero. Nonetheless, we can define an invariant, named the spin Chern number for this state
\begin{equation}~\label{eq:spinchern}
C_{sp}=|C_\uparrow-C_\downarrow|, 
\end{equation}
where $C_\uparrow$ ($C_\downarrow$) is the first Chern number associated with the $\uparrow$ ($\downarrow$) spin component. Notice that $C_{sp}$ is nontrivial and $C_{sp}=2$ in the entire topological regime ($|m_0/t_0|<2$) and $C_{sp}=0$ for $|m_0/t_0|>2$. See Fig.~\ref{fig:pwavespintsc}(a). However, we cannot apply an external magnetic field to probe nontrivial $C_{sp}$, as it breaks the TRS, which is a conserved symmetry for the class DIII. Hence, the only meaningful experimentally measurable transport response of the $p \pm ip$ pairing is the longitudinal thermal conductance ($G^{th}_{xx}$). A six-terminal setup is employed to compute $G^{th}_{xx}$, as shown in Fig.~\ref{fig:pwavespintsc}(b). The computational procedure is identical to the one described in details in Sec.~\ref{sec:pipcleanthermal} (see also Appendix~\ref{append:scattering}). From the longitudinal thermal resistance $R^{th}_{xx}=(T_3-T_2)/{I_{th}}$, we compute $G^{th}_{xx}=\left( R^{th}_{xx} \right)^{-1}$ and find that 
\begin{equation}
G^{th}_{xx}= \frac{C_{sp}}{2} \kappa_0,
\end{equation}
where the factor of $1/2$ compensates the Nambu doubling. The results are shown in Fig.~\ref{fig:pwavespintsc}(c). The quantized $G^{th}_{xx}$ in the clean $p \pm ip$ paired state results from two counter-propagating Majorana edge modes for opposite spin projections living on the same edge of the system, which are shown in Fig.~\ref{fig:pwavespintsc}(b). We numerically confirmed that $\kappa_{xy}=0$ for the $p \pm ip$ paired state, as it supports two counter-propagating edge modes on each edge.

To examine the robustness of the quantized longitudinal thermal conductance against random charge impurities, we compute its disorder averaged values $\langle G^{th}_{xx} \rangle$ by adding a term $V(\vec{r})\Gamma_3$ to each site of the scattering region. Here as well $V(\vec{r})$ is uniformly and randomly distributed in the range $[-W/2,W/2]$ on every site of the scattering region, and $W$ denotes the disorder strength. We observe that $\langle G_{xx}^{th} \rangle$ retains its quantized value (in units of $\kappa_0$) up to a moderate disorder strength, as shown in Fig.~\ref{fig:pwavespintsc}(d), beyond which it acquires nonuniversal and nonquantized values before vanishing at sufficiently strong disorder. The quantization of $\langle G_{xx}^{th} \rangle$ in the weak disorder regime results from the absence of any backscattering between two counter-propagating Majorana edge modes of opposite spin projections in the $p \pm ip$ paired state, living on the same edge. As these edge modes have opposite spin projections, any backscattering between them is forbidden by any time-reversal symmetric disorder (such as the random charge impurities), which is the symmetry of class DIII (see Table~\ref{table:symmetryclassification}). The values of $\langle G_{xx}^{th} \rangle$ typically saturate after averaging over 150 independent disorder realizations, irrespective of its strength. See Fig.~\ref{fig:pwavespintsc}(e). In Fig.~\ref{fig:pwavespintsc}(d) and (e), we also display the standard deviation of $G_{xx}^{th}$ for each value of $W$, which also saturates after averaging over 150 independent disorder realizations. The qualitative behavior of the standard deviation of $G_{xx}^{th}$ is similar to the ones we discussed for all the previously cases.


\section{Summary and discussions}~\label{sec:conclusions}

To summarize, here we numerically compute the (half-)quantized thermal ($\kappa_{xy}$) and spin ($\sigma^{sp}_{xy}$) Hall responses, as well as the quantized longitudinal thermal conductance ($G^{th}_{xx}$) of prominent gapped two-dimensional topological paired states from different AZ symmetry classes [Table.~\ref{table:symmetryclassification}] by employing scattering matrix formalism using the Kwant software package on finite lattice regularized mesoscopic systems. The transverse thermal Hall conductivity and longitudinal thermal conductance are computed at sufficiently low temperatures ($T=0.01$), and their (half-)quantization values are quoted in units of $\kappa_0=\pi^2 k^2_B T/(3h)$. On the other hand, the zero temperature spin Hall conductance is reported in units of $\sigma^{sp}_0 =\hbar/(8\pi)$. These quantities in clean systems are shown to be tied with the bulk topological invariants of the corresponding effective single-particle BdG Hamiltonian, which continue to feature robustness in the presence of weak random charge impurities, manifesting the stability of the bulk topology in the weak disorder regime. In particular, weakly disordered class D $p+ip$ and class DIII $p \pm ip$ paired states respectively display $\langle \kappa_{xy} \rangle =\pm \kappa_0/2$ [Fig.~\ref{fig:pwavetsc}] and $\langle G^{th}_{xx} \rangle=\kappa_0$ [Fig.~\ref{fig:pwavespintsc}]. Finally, class C spin-singlet $d+id$ state supports both quantized $\langle \kappa_{xy} \rangle$ [Fig.~\ref{fig:dwavethermaltsc}] and $\langle \sigma^{sp}_{xy} \rangle$ [Fig.~\ref{fig:dwavespintsc}], and their ratio defines the modified Lorentz number 
\begin{equation}
L_m=\frac{\underset{T \to 0}{\lim} \; \left( \langle \kappa_{xy} \rangle /\kappa_0 \right)}{\langle \sigma^{sp}_{xy}\rangle /\sigma^{sp}_0}=1,
\end{equation}
which remains pinned at this universal value of \emph{unity} in the weak disorder regime. By contrast, in the strong disorder regime all these topological responses disappear, indicating onset of trivial thermal insulators. Vanishing topological transport responses, such as $\kappa_{xy}$, $\sigma^{sp}_{xy}$ and $G^{th}_{xx}$, for TSCs from different symmetry classes (class D, class C and class DIII) in the strong disorder regime is rooted in a universal mechanism. Note that the localization length of all the edge modes $\xi \sim (\Delta^{\rm bulk}_{\rm gap})^{-1}$, where $\Delta^{\rm bulk}_{\rm gap}$ is the bulk gap of the TSC. As the disorder strength ($W$) is increased, $\Delta^{\rm bulk}_{\rm gap}$ decreases, and thus $\xi$ increases, causing hybridization between the edge modes of same spin projection but living on the \emph{opposite} edges of the system. In the strong disorder regime, such a hybridization becomes sufficiently strong and the edge modes become \emph{gapped}. Then all the topological responses disappear from the system, which we find for TSCs, belonging to any symmetry class. Altogether, here we develop concrete numerical methodologies for the computation of various quantized thermal and spin responses in the clean and the dirty topological superconductors, encompassing all three allowed AZ symmetry classes in two dimensions. Recent success in the experimental measurements of the thermal Hall conductivity in spin liquids, integer and fractional quantum Hall states in six-terminal Hall bar geometry~\cite{Banerjee2018a, Banerjee2018b, Kasahara2018, Srivastav2019, Breton2022} should make this analysis pertinent in real materials for which the effective model Hamiltonian can be constructed from lattice-based symmetry constraints. On such available model Hamiltonian our methodology can be directly applied, which is available on Zenodo~\cite{zenodoTSCSanjib}. In future, it will be worthwhile extending the notion of thermal and spin responses to \emph{noncrystalline} topological superconductors on fractals, amorphous materials and quasicrystals~\cite{MannaDas2022}, for example.

We note that the disorder averaged low temperature thermal Hall conductivity $\langle \kappa_{xy} \rangle$ for the $p+ip$ and $d+id$ paired states, and the longitudinal thermal conductance $\langle G^{th}_{xx} \rangle$ of the $p \pm ip$ superconductor initially increase from their (half-)quantized values at moderate disorder strength before they all decay to zero at sufficiently strong disorder. Their quoted disorder averaged values are saturated with respect to the number of independent disorder realizations ($n$). See panel (d) and (e) of Figs.~\ref{fig:pwavetsc},~\ref{fig:dwavethermaltsc} and~\ref{fig:pwavespintsc}. This feature is, however, extremely small for the zero temperature spin Hall conductivity for the $d+id$ paired state. See Fig.~\ref{fig:dwavespintsc}(c) and (d). The microscopic origin of such a peculiar observation is presently not clear. We suspect that at moderate disorder the system fragments into multiple \emph{thermally excited} islands or droplets of topological and trivial paired states, each interface between them within the scattering region gives rise to Majorana edge modes with weak hybridization among them, yielding  enhanced, but nonquantized values of $\langle \kappa_{xy} \rangle$ and $\langle G^{th}_{xx} \rangle$. In the strong disorder limit, the number of such islands possibly increases, causing a strong hybridization among a large number of such Majorana edge modes living within the system, leading to vanishing $\langle \kappa_{xy} \rangle$ and $\langle G^{th}_{xx} \rangle$. Such a scenario can be tested by computing the local topological markers of disorder topological superconductors, which can in principle be extracted for any AZ symmetry class~\cite{Resta2011, WeiChen2023}. We expect that the local topological marker can reveal such droplet structure in disordered topological superconductors at finite temperature (if exists at all). We leave this open question as a topic for a future investigation.

One of the practical challenges in the field of planar topological superconductivity involves the identification of real quantum materials that can harbor such exotic quantum phase of matter at low temperatures~\cite{FuAndo2015, Sato2017, Mandal2023}, with, however, Sr$_2$RuO$_4$ standing as one prominent candidate~\cite{MackenzieMaeno2003, Kallin2012, Maenoetal2012}, for example. Nonetheless, over the past several years quantum crystals potentially fostering topological insulators emerged as promising materials where on-site or local or momentum-independent paired states can nucleate at low temperature and represent topological superconductors, especially when these materials are doped or chemically substituted or intercalated to sustain an underlying Fermi surface, conducive for the Cooper pairing. As such, local pairings in these family of materials inherit topology from the normal state band structure of charged fermions~\cite{Das2023, FuBerg2010, LiangFu2014, BRoy2020, RoyJuricic2021, MandalRoy2023}. This proposal has received promising supports in three-dimensional (3D) intercalated and doped TIs, namely Cu$_x$Bi$_2$Se$_3$ and Sn$_{1-x}$In$_x$Te, that become superconductors at low temperatures with a few Kelvin transition temperature, featuring surface zero-bias-conductance peak, resulting from gapless surface Majorana fermions, suggesting topological nature of the underlying paired state~\cite{Ando2011PRL, Ando2012PRL, Ando2013PRB}. In addition, proximity effect can be an efficient, realistic and experimentally viable route to induce topological superconductivity in various two-dimensional doped topological insulator materials. In this case, the parent superconductor is typically a fully gapped and trivial $s$-wave one (such as Nb) with no gapless zero-energy excitation. Thus, its presence do not affect the topological thermal and spin responses of the proximity-induced TSCs, solely resulting from their zero-energy fermionic BdG quasiparticles living at the edges (topological edge modes).

Experimental observations of possible 3D TSCs in doped and intercalated 3D TIs, make it a promising avenue to harness 2D TSCs on similar material platforms. In particular, doped or proximetized quantum anomalous Hall insulators with already broken TRS in the normal state (class A) can be an ideal place to harness a $p+ip$ paired state~\cite{Das2023}. Recently realized quantum anomalous Hall insulators in magnetically doped (by Cr or V or Fe, for example) thin layer of three-dimensional topological insulators, such as Bi$_2$Se$_3$, Bi$_2$Te$_3$ and Sb$_2$Te$_3$~\cite{ZhongFang2010, QKXu2013, Chang2015}, are, therefore, promising in this respect. In the same spirit, doped or proximetized TRS preserving quantum spin Hall insulators (class AII), such as CdTe-HgTe~\cite{BHZ2006, Konig2007} and InAs-SbTe~\cite{Knez2011} quantum wells, constitute to a suitable ground to realize a $p \pm ip$ paired state~\cite{BRoy2020}. Finally, high-$T_c$ cuprate superconductors (in particular, Bi$_2$Sr$_2$CaCu$_2$O$_8$) are promising to realize a $d+id$ paired state~\cite{Krishana1997, Movshovich1998, Laughlin1998}. Although, a clear signature of the $d+id$ paired state in Bi$_2$Sr$_2$CaCu$_2$O$_8$ thus far remains illusive, a \emph{twisted} interface between bilayer Bi$_2$Sr$_2$CaCu$_2$O$_{8+x}$ can host such exotic paired state near $45^\circ$ twist angle~\cite{Pkim2021}.

Our direct computation of the quantized longitudinal thermal conductance $G^{th}_{xx}$ for the class DIII $p \pm ip$ paired state with a net zero first Chern number ($C$) can have far reaching consequences. For example, if it happens that inside a two-dimensional topological paired state the total first Chern number is zero, which may occur due to the presence of spin or other (such as orbital and/or sublattice) degrees of freedom or multiple Fermi pockets around which the net first Chern number cancels out, still $G^{th}_{xx}$ can probe the total number of topologically robust Majorana edge modes in the superconducting ground state, each of which contributes $\kappa_0/2$ to $G^{th}_{xx}$. In the same spirit, $G^{th}_{xx}$ can also probe \emph{weak} planar topological superconductors, devoid of any strong topological invariant~\cite{Das2023}. Such a scenario may appear commonly in the superconducting ground state of doped crystalline topological materials~\cite{Fu2011, Slager2012, Shiozaki2014}, typically featuring band inversion at an even number of high symmetry points in the BZ connected via crystal symmetries, that are nowadays routinely found in nature by employing topological quantum chemistry~\cite{bernevig2017, Vishwanath2017NatComm, ashvin2018, Zhang2019, Vergniory2019, Tang2019}. These avenues will be explored systematically in the future.

\acknowledgments 

S.K.D.\ was supported by a Startup Grant of B.R.\ from Lehigh University. B.R.\ was supported by NSF CAREER Grant No. DMR-2238679. Portions of this research were conducted on Lehigh University's Research Computing infrastructure partially supported by NSF Award No.~2019035.

\appendix

\section{Details of scattering matrix}~\label{append:scattering}

In this section, we present the details of the scattering matrix formalism, employed to compute the thermal and spin transport properties of 2D TSCs. The rectangular scattering region (system) is attached to six leads. All the leads are semi-infinite and they supply fermions to the scattering region. A thermal/spin current flows between the leads in the horizonal/longitudinal direction (Lead 1 and Lead 4), which generates a temperature/magnetization drop in the transverse leads (Lead 2, Lead 3, Lead 5, Lead 6). From such a temperature/magnetization drop, generated between different leads, we compute the thermal and spin transport quantities, described below.

The requisite scattering matrix ($\mathcal{S}$) is obtained by solving the linear equation $\Psi_{\rm out} = \mathcal{S}\Psi_{\rm in}$, where 
\begin{equation}
\mathcal{S} = \begin{pmatrix}
r & t'\\
t & r'
\end{pmatrix},
\end{equation}
and $r$ and $t$ are the reflection and transmission blocks of the scattering matrix, respectively, with $|r|^2+|t|^2=1$ preserving the unitarity of the scattering matrix. Here, $\Psi_{\rm in}$ ($\Psi_{\rm out}$) is the incoming (outgoing) wavefunction, entering (leaving) the scattering region.

For the computation of the THC ($\kappa_{xy}$), a thermal current ($I_{th}$) is flows between Leads 1 and Lead 4. Then $\kappa_{xy}$ is obtained from temperature drop of the transverse leads, computed from the linear relation between the thermal current ($\mathbf{I}_{th}$) and temperature ($\mathbf{T}$), discussed in Sec.~\ref{sec:pipcleanthermal}.

In the same setup, for the computation of the spin Hall conductivity ($\sigma_{xy}^{sp}$), a spin current flows between Lead 1 and Lead 4, which results in a magnetization drop in the vertical leads. In this case, a linear relation between spin current ($\mathbf{I}_{sp}$) and magnetization ($\mathbf{M}$) is used to extract $\sigma_{xy}^{sp}$. See Sec.~\ref{subsec:dwavespinhall}.

Finally, with the same setup, longitudinal thermal conductance ($G_{xx}^{th}$) is obtained once the temperature drop between Lead 2 and Lead 3 is taken into account. In this case, the computation scheme remains same as that of THC, except now we take the temperature drop between the longitudinal leads (Lead 2 and Lead 3).

\bibliography{ref}

\begin{thebibliography}{82}%
\makeatletter
\providecommand \@ifxundefined [1]{%
 \@ifx{#1\undefined}
}%
\providecommand \@ifnum [1]{%
 \ifnum #1\expandafter \@firstoftwo
 \else \expandafter \@secondoftwo
 \fi
}%
\providecommand \@ifx [1]{%
 \ifx #1\expandafter \@firstoftwo
 \else \expandafter \@secondoftwo
 \fi
}%
\providecommand \natexlab [1]{#1}%
\providecommand \enquote  [1]{``#1''}%
\providecommand \bibnamefont  [1]{#1}%
\providecommand \bibfnamefont [1]{#1}%
\providecommand \citenamefont [1]{#1}%
\providecommand \href@noop [0]{\@secondoftwo}%
\providecommand \href [0]{\begingroup \@sanitize@url \@href}%
\providecommand \@href[1]{\@@startlink{#1}\@@href}%
\providecommand \@@href[1]{\endgroup#1\@@endlink}%
\providecommand \@sanitize@url [0]{\catcode `\\12\catcode `\$12\catcode
  `\&12\catcode `\#12\catcode `\^12\catcode `\_12\catcode `\%12\relax}%
\providecommand \@@startlink[1]{}%
\providecommand \@@endlink[0]{}%
\providecommand \url  [0]{\begingroup\@sanitize@url \@url }%
\providecommand \@url [1]{\endgroup\@href {#1}{\urlprefix }}%
\providecommand \urlprefix  [0]{URL }%
\providecommand \Eprint [0]{\href }%
\providecommand \doibase [0]{https://doi.org/}%
\providecommand \selectlanguage [0]{\@gobble}%
\providecommand \bibinfo  [0]{\@secondoftwo}%
\providecommand \bibfield  [0]{\@secondoftwo}%
\providecommand \translation [1]{[#1]}%
\providecommand \BibitemOpen [0]{}%
\providecommand \bibitemStop [0]{}%
\providecommand \bibitemNoStop [0]{.\EOS\space}%
\providecommand \EOS [0]{\spacefactor3000\relax}%
\providecommand \BibitemShut  [1]{\csname bibitem#1\endcsname}%
\let\auto@bib@innerbib\@empty
\bibitem [{\citenamefont {Hasan}\ and\ \citenamefont {Kane}(2010)}]{Hasan2010}%
  \BibitemOpen
  \bibfield  {author} {\bibinfo {author} {\bibfnamefont {M.~Z.}\ \bibnamefont
  {Hasan}}\ and\ \bibinfo {author} {\bibfnamefont {C.~L.}\ \bibnamefont
  {Kane}},\ }\bibfield  {title} {\bibinfo {title} {{Colloquium: Topological
  insulators}},\ }\href {https://doi.org/10.1103/RevModPhys.82.3045} {\bibfield
   {journal} {\bibinfo  {journal} {Rev. Mod. Phys.}\ }\textbf {\bibinfo
  {volume} {82}},\ \bibinfo {pages} {3045} (\bibinfo {year}
  {2010})}\BibitemShut {NoStop}%
\bibitem [{\citenamefont {Qi}\ and\ \citenamefont {Zhang}(2011)}]{Qi2011}%
  \BibitemOpen
  \bibfield  {author} {\bibinfo {author} {\bibfnamefont {X.-L.}\ \bibnamefont
  {Qi}}\ and\ \bibinfo {author} {\bibfnamefont {S.-C.}\ \bibnamefont {Zhang}},\
  }\bibfield  {title} {\bibinfo {title} {{Topological insulators and
  superconductors}},\ }\href {https://doi.org/10.1103/RevModPhys.83.1057}
  {\bibfield  {journal} {\bibinfo  {journal} {Rev. Mod. Phys.}\ }\textbf
  {\bibinfo {volume} {83}},\ \bibinfo {pages} {1057} (\bibinfo {year}
  {2011})}\BibitemShut {NoStop}%
\bibitem [{\citenamefont {Altland}\ and\ \citenamefont
  {Zirnbauer}(1997)}]{Altland1997}%
  \BibitemOpen
  \bibfield  {author} {\bibinfo {author} {\bibfnamefont {A.}~\bibnamefont
  {Altland}}\ and\ \bibinfo {author} {\bibfnamefont {M.~R.}\ \bibnamefont
  {Zirnbauer}},\ }\bibfield  {title} {\bibinfo {title} {{Nonstandard symmetry
  classes in mesoscopic normal-superconducting hybrid structures}},\ }\href
  {https://doi.org/10.1103/PhysRevB.55.1142} {\bibfield  {journal} {\bibinfo
  {journal} {Phys. Rev. B}\ }\textbf {\bibinfo {volume} {55}},\ \bibinfo
  {pages} {1142} (\bibinfo {year} {1997})}\BibitemShut {NoStop}%
\bibitem [{\citenamefont {Kane}\ and\ \citenamefont
  {Mele}(2005)}]{kanemele2006}%
  \BibitemOpen
  \bibfield  {author} {\bibinfo {author} {\bibfnamefont {C.~L.}\ \bibnamefont
  {Kane}}\ and\ \bibinfo {author} {\bibfnamefont {E.~J.}\ \bibnamefont
  {Mele}},\ }\bibfield  {title} {\bibinfo {title} {{${Z}_{2}$ Topological Order
  and the Quantum Spin Hall Effect}},\ }\href
  {https://doi.org/10.1103/PhysRevLett.95.146802} {\bibfield  {journal}
  {\bibinfo  {journal} {Phys. Rev. Lett.}\ }\textbf {\bibinfo {volume} {95}},\
  \bibinfo {pages} {146802} (\bibinfo {year} {2005})}\BibitemShut {NoStop}%
\bibitem [{\citenamefont {{Bernevig}}\ \emph {et~al.}(2006)\citenamefont
  {{Bernevig}}, \citenamefont {{Hughes}},\ and\ \citenamefont
  {{Zhang}}}]{BHZ2006}%
  \BibitemOpen
  \bibfield  {author} {\bibinfo {author} {\bibfnamefont {B.~A.}\ \bibnamefont
  {{Bernevig}}}, \bibinfo {author} {\bibfnamefont {T.~L.}\ \bibnamefont
  {{Hughes}}},\ and\ \bibinfo {author} {\bibfnamefont {S.-C.}\ \bibnamefont
  {{Zhang}}},\ }\bibfield  {title} {\bibinfo {title} {{Quantum Spin Hall Effect
  and Topological Phase Transition in HgTe Quantum Wells}},\ }\href
  {https://doi.org/10.1126/science.1133734} {\bibfield  {journal} {\bibinfo
  {journal} {Science}\ }\textbf {\bibinfo {volume} {314}},\ \bibinfo {pages}
  {1757} (\bibinfo {year} {2006})}\BibitemShut {NoStop}%
\bibitem [{\citenamefont {Fu}\ \emph {et~al.}(2007)\citenamefont {Fu},
  \citenamefont {Kane},\ and\ \citenamefont {Mele}}]{FuKaneMele2007}%
  \BibitemOpen
  \bibfield  {author} {\bibinfo {author} {\bibfnamefont {L.}~\bibnamefont
  {Fu}}, \bibinfo {author} {\bibfnamefont {C.~L.}\ \bibnamefont {Kane}},\ and\
  \bibinfo {author} {\bibfnamefont {E.~J.}\ \bibnamefont {Mele}},\ }\bibfield
  {title} {\bibinfo {title} {{Topological Insulators in Three Dimensions}},\
  }\href {https://doi.org/10.1103/PhysRevLett.98.106803} {\bibfield  {journal}
  {\bibinfo  {journal} {Phys. Rev. Lett.}\ }\textbf {\bibinfo {volume} {98}},\
  \bibinfo {pages} {106803} (\bibinfo {year} {2007})}\BibitemShut {NoStop}%
\bibitem [{\citenamefont {Fu}\ and\ \citenamefont {Kane}(2007)}]{Fukane2007}%
  \BibitemOpen
  \bibfield  {author} {\bibinfo {author} {\bibfnamefont {L.}~\bibnamefont
  {Fu}}\ and\ \bibinfo {author} {\bibfnamefont {C.~L.}\ \bibnamefont {Kane}},\
  }\bibfield  {title} {\bibinfo {title} {{Topological insulators with inversion
  symmetry}},\ }\href {https://doi.org/10.1103/PhysRevB.76.045302} {\bibfield
  {journal} {\bibinfo  {journal} {Phys. Rev. B}\ }\textbf {\bibinfo {volume}
  {76}},\ \bibinfo {pages} {045302} (\bibinfo {year} {2007})}\BibitemShut
  {NoStop}%
\bibitem [{\citenamefont {Moore}\ and\ \citenamefont
  {Balents}(2007)}]{moorebalents2007}%
  \BibitemOpen
  \bibfield  {author} {\bibinfo {author} {\bibfnamefont {J.~E.}\ \bibnamefont
  {Moore}}\ and\ \bibinfo {author} {\bibfnamefont {L.}~\bibnamefont
  {Balents}},\ }\bibfield  {title} {\bibinfo {title} {{Topological invariants
  of time-reversal-invariant band structures}},\ }\href
  {https://doi.org/10.1103/PhysRevB.75.121306} {\bibfield  {journal} {\bibinfo
  {journal} {Phys. Rev. B}\ }\textbf {\bibinfo {volume} {75}},\ \bibinfo
  {pages} {121306} (\bibinfo {year} {2007})}\BibitemShut {NoStop}%
\bibitem [{\citenamefont {Schnyder}\ \emph {et~al.}(2008)\citenamefont
  {Schnyder}, \citenamefont {Ryu}, \citenamefont {Furusaki},\ and\
  \citenamefont {Ludwig}}]{Schnyder2008}%
  \BibitemOpen
  \bibfield  {author} {\bibinfo {author} {\bibfnamefont {A.~P.}\ \bibnamefont
  {Schnyder}}, \bibinfo {author} {\bibfnamefont {S.}~\bibnamefont {Ryu}},
  \bibinfo {author} {\bibfnamefont {A.}~\bibnamefont {Furusaki}},\ and\
  \bibinfo {author} {\bibfnamefont {A.~W.~W.}\ \bibnamefont {Ludwig}},\
  }\bibfield  {title} {\bibinfo {title} {{Classification of topological
  insulators and superconductors in three spatial dimensions}},\ }\href
  {https://doi.org/10.1103/PhysRevB.78.195125} {\bibfield  {journal} {\bibinfo
  {journal} {Phys. Rev. B}\ }\textbf {\bibinfo {volume} {78}},\ \bibinfo
  {pages} {195125} (\bibinfo {year} {2008})}\BibitemShut {NoStop}%
\bibitem [{\citenamefont {Kitaev}(2009)}]{Kitaev2009}%
  \BibitemOpen
  \bibfield  {author} {\bibinfo {author} {\bibfnamefont {A.}~\bibnamefont
  {Kitaev}},\ }\bibfield  {title} {\bibinfo {title} {{Periodic table for
  topological insulators and superconductors}},\ }\href
  {https://doi.org/10.1063/1.3149495} {\bibfield  {journal} {\bibinfo
  {journal} {AIP Conf. Proc.}\ }\textbf {\bibinfo {volume} {1134}},\ \bibinfo
  {pages} {22} (\bibinfo {year} {2009})}\BibitemShut {NoStop}%
\bibitem [{\citenamefont {Roy}(2009)}]{rahulroy2009}%
  \BibitemOpen
  \bibfield  {author} {\bibinfo {author} {\bibfnamefont {R.}~\bibnamefont
  {Roy}},\ }\bibfield  {title} {\bibinfo {title} {{Topological phases and the
  quantum spin Hall effect in three dimensions}},\ }\href
  {https://doi.org/10.1103/PhysRevB.79.195322} {\bibfield  {journal} {\bibinfo
  {journal} {Phys. Rev. B}\ }\textbf {\bibinfo {volume} {79}},\ \bibinfo
  {pages} {195322} (\bibinfo {year} {2009})}\BibitemShut {NoStop}%
\bibitem [{\citenamefont {Ryu}\ \emph {et~al.}(2010)\citenamefont {Ryu},
  \citenamefont {Schnyder}, \citenamefont {Furusaki},\ and\ \citenamefont
  {Ludwig}}]{Ryu2010}%
  \BibitemOpen
  \bibfield  {author} {\bibinfo {author} {\bibfnamefont {S.}~\bibnamefont
  {Ryu}}, \bibinfo {author} {\bibfnamefont {A.~P.}\ \bibnamefont {Schnyder}},
  \bibinfo {author} {\bibfnamefont {A.}~\bibnamefont {Furusaki}},\ and\
  \bibinfo {author} {\bibfnamefont {A.~W.~W.}\ \bibnamefont {Ludwig}},\
  }\bibfield  {title} {\bibinfo {title} {{Topological insulators and
  superconductors: tenfold way and dimensional hierarchy}},\ }\href
  {https://doi.org/10.1088/1367-2630/12/6/065010} {\bibfield  {journal}
  {\bibinfo  {journal} {New J. Phys.}\ }\textbf {\bibinfo {volume} {12}},\
  \bibinfo {pages} {065010} (\bibinfo {year} {2010})}\BibitemShut {NoStop}%
\bibitem [{\citenamefont {Soluyanov}\ and\ \citenamefont
  {Vanderbilt}(2011)}]{vanderbilt2011}%
  \BibitemOpen
  \bibfield  {author} {\bibinfo {author} {\bibfnamefont {A.~A.}\ \bibnamefont
  {Soluyanov}}\ and\ \bibinfo {author} {\bibfnamefont {D.}~\bibnamefont
  {Vanderbilt}},\ }\bibfield  {title} {\bibinfo {title} {{Computing topological
  invariants without inversion symmetry}},\ }\href
  {https://doi.org/10.1103/PhysRevB.83.235401} {\bibfield  {journal} {\bibinfo
  {journal} {Phys. Rev. B}\ }\textbf {\bibinfo {volume} {83}},\ \bibinfo
  {pages} {235401} (\bibinfo {year} {2011})}\BibitemShut {NoStop}%
\bibitem [{\citenamefont {Chiu}\ \emph {et~al.}(2016)\citenamefont {Chiu},
  \citenamefont {Teo}, \citenamefont {Schnyder},\ and\ \citenamefont
  {Ryu}}]{Chiu2016}%
  \BibitemOpen
  \bibfield  {author} {\bibinfo {author} {\bibfnamefont {C.-K.}\ \bibnamefont
  {Chiu}}, \bibinfo {author} {\bibfnamefont {J.~C.~Y.}\ \bibnamefont {Teo}},
  \bibinfo {author} {\bibfnamefont {A.~P.}\ \bibnamefont {Schnyder}},\ and\
  \bibinfo {author} {\bibfnamefont {S.}~\bibnamefont {Ryu}},\ }\bibfield
  {title} {\bibinfo {title} {{Classification of topological quantum matter with
  symmetries}},\ }\href {https://doi.org/10.1103/RevModPhys.88.035005}
  {\bibfield  {journal} {\bibinfo  {journal} {Rev. Mod. Phys.}\ }\textbf
  {\bibinfo {volume} {88}},\ \bibinfo {pages} {035005} (\bibinfo {year}
  {2016})}\BibitemShut {NoStop}%
\bibitem [{\citenamefont {Qi}\ \emph {et~al.}(2008)\citenamefont {Qi},
  \citenamefont {Hughes},\ and\ \citenamefont {Zhang}}]{SCZhang2008}%
  \BibitemOpen
  \bibfield  {author} {\bibinfo {author} {\bibfnamefont {X.-L.}\ \bibnamefont
  {Qi}}, \bibinfo {author} {\bibfnamefont {T.~L.}\ \bibnamefont {Hughes}},\
  and\ \bibinfo {author} {\bibfnamefont {S.-C.}\ \bibnamefont {Zhang}},\
  }\bibfield  {title} {\bibinfo {title} {{Topological field theory of
  time-reversal invariant insulators}},\ }\href
  {https://doi.org/10.1103/PhysRevB.78.195424} {\bibfield  {journal} {\bibinfo
  {journal} {Phys. Rev. B}\ }\textbf {\bibinfo {volume} {78}},\ \bibinfo
  {pages} {195424} (\bibinfo {year} {2008})}\BibitemShut {NoStop}%
\bibitem [{\citenamefont {Dzero}\ \emph {et~al.}(2010)\citenamefont {Dzero},
  \citenamefont {Sun}, \citenamefont {Galitski},\ and\ \citenamefont
  {Coleman}}]{Dzero2010}%
  \BibitemOpen
  \bibfield  {author} {\bibinfo {author} {\bibfnamefont {M.}~\bibnamefont
  {Dzero}}, \bibinfo {author} {\bibfnamefont {K.}~\bibnamefont {Sun}}, \bibinfo
  {author} {\bibfnamefont {V.}~\bibnamefont {Galitski}},\ and\ \bibinfo
  {author} {\bibfnamefont {P.}~\bibnamefont {Coleman}},\ }\bibfield  {title}
  {\bibinfo {title} {{Topological Kondo Insulators}},\ }\href
  {https://doi.org/10.1103/PhysRevLett.104.106408} {\bibfield  {journal}
  {\bibinfo  {journal} {Phys. Rev. Lett.}\ }\textbf {\bibinfo {volume} {104}},\
  \bibinfo {pages} {106408} (\bibinfo {year} {2010})}\BibitemShut {NoStop}%
\bibitem [{\citenamefont {Roy}\ \emph {et~al.}(2014)\citenamefont {Roy},
  \citenamefont {Sau}, \citenamefont {Dzero},\ and\ \citenamefont
  {Galitski}}]{RoyKondo2014}%
  \BibitemOpen
  \bibfield  {author} {\bibinfo {author} {\bibfnamefont {B.}~\bibnamefont
  {Roy}}, \bibinfo {author} {\bibfnamefont {J.~D.}\ \bibnamefont {Sau}},
  \bibinfo {author} {\bibfnamefont {M.}~\bibnamefont {Dzero}},\ and\ \bibinfo
  {author} {\bibfnamefont {V.}~\bibnamefont {Galitski}},\ }\bibfield  {title}
  {\bibinfo {title} {{Surface theory of a family of topological Kondo
  insulators}},\ }\href {https://doi.org/10.1103/PhysRevB.90.155314} {\bibfield
   {journal} {\bibinfo  {journal} {Phys. Rev. B}\ }\textbf {\bibinfo {volume}
  {90}},\ \bibinfo {pages} {155314} (\bibinfo {year} {2014})}\BibitemShut
  {NoStop}%
\bibitem [{\citenamefont {{Dzero}}\ \emph {et~al.}(2016)\citenamefont
  {{Dzero}}, \citenamefont {{Xia}}, \citenamefont {{Galitski}},\ and\
  \citenamefont {{Coleman}}}]{DzeroReview2016}%
  \BibitemOpen
  \bibfield  {author} {\bibinfo {author} {\bibfnamefont {M.}~\bibnamefont
  {{Dzero}}}, \bibinfo {author} {\bibfnamefont {J.}~\bibnamefont {{Xia}}},
  \bibinfo {author} {\bibfnamefont {V.}~\bibnamefont {{Galitski}}},\ and\
  \bibinfo {author} {\bibfnamefont {P.}~\bibnamefont {{Coleman}}},\ }\bibfield
  {title} {\bibinfo {title} {{Topological Kondo Insulators}},\ }\href
  {https://doi.org/10.1146/annurev-conmatphys-031214-014749} {\bibfield
  {journal} {\bibinfo  {journal} {Annu. Rev. Condens. Matter Phys.}\ }\textbf
  {\bibinfo {volume} {7}},\ \bibinfo {pages} {249} (\bibinfo {year}
  {2016})}\BibitemShut {NoStop}%
\bibitem [{\citenamefont {Volovik}(2009)}]{Volovik2009}%
  \BibitemOpen
  \bibfield  {author} {\bibinfo {author} {\bibfnamefont {G.~E.}\ \bibnamefont
  {Volovik}},\ }\href
  {https://doi.org/10.1093/acprof:oso/9780199564842.001.0001} {\emph {\bibinfo
  {title} {{The Universe in a Helium Droplet}}}}\ (\bibinfo  {publisher}
  {Oxford University Press, Oxford, UK},\ \bibinfo {year} {2009})\BibitemShut
  {NoStop}%
\bibitem [{\citenamefont {Read}\ and\ \citenamefont {Green}(2000)}]{Read2000}%
  \BibitemOpen
  \bibfield  {author} {\bibinfo {author} {\bibfnamefont {N.}~\bibnamefont
  {Read}}\ and\ \bibinfo {author} {\bibfnamefont {D.}~\bibnamefont {Green}},\
  }\bibfield  {title} {\bibinfo {title} {{Paired states of fermions in two
  dimensions with breaking of parity and time-reversal symmetries and the
  fractional quantum {Hall} effect}},\ }\href
  {https://doi.org/10.1103/PhysRevB.61.10267} {\bibfield  {journal} {\bibinfo
  {journal} {Phys. Rev. B}\ }\textbf {\bibinfo {volume} {61}},\ \bibinfo
  {pages} {10267} (\bibinfo {year} {2000})}\BibitemShut {NoStop}%
\bibitem [{\citenamefont {{Kitaev}}(2006)}]{Kitaev2006}%
  \BibitemOpen
  \bibfield  {author} {\bibinfo {author} {\bibfnamefont {A.}~\bibnamefont
  {{Kitaev}}},\ }\bibfield  {title} {\bibinfo {title} {{Anyons in an exactly
  solved model and beyond}},\ }\href
  {https://doi.org/10.1016/j.aop.2005.10.005} {\bibfield  {journal} {\bibinfo
  {journal} {Ann. Phys.}\ }\textbf {\bibinfo {volume} {321}},\ \bibinfo {pages}
  {2} (\bibinfo {year} {2006})}\BibitemShut {NoStop}%
\bibitem [{\citenamefont {Wang}\ \emph {et~al.}(2011)\citenamefont {Wang},
  \citenamefont {Qi},\ and\ \citenamefont {Zhang}}]{SCZhang2011PRB}%
  \BibitemOpen
  \bibfield  {author} {\bibinfo {author} {\bibfnamefont {Z.}~\bibnamefont
  {Wang}}, \bibinfo {author} {\bibfnamefont {X.-L.}\ \bibnamefont {Qi}},\ and\
  \bibinfo {author} {\bibfnamefont {S.-C.}\ \bibnamefont {Zhang}},\ }\bibfield
  {title} {\bibinfo {title} {{Topological field theory and thermal responses of
  interacting topological superconductors}},\ }\href
  {https://doi.org/10.1103/PhysRevB.84.014527} {\bibfield  {journal} {\bibinfo
  {journal} {Phys. Rev. B}\ }\textbf {\bibinfo {volume} {84}},\ \bibinfo
  {pages} {014527} (\bibinfo {year} {2011})}\BibitemShut {NoStop}%
\bibitem [{\citenamefont {Sumiyoshi}\ and\ \citenamefont
  {Fujimoto}(2013)}]{Fujimoto2013}%
  \BibitemOpen
  \bibfield  {author} {\bibinfo {author} {\bibfnamefont {H.}~\bibnamefont
  {Sumiyoshi}}\ and\ \bibinfo {author} {\bibfnamefont {S.}~\bibnamefont
  {Fujimoto}},\ }\bibfield  {title} {\bibinfo {title} {{Quantum Thermal Hall
  Effect in a Time-Reversal-Symmetry-Broken Topological Superconductor in Two
  Dimensions: Approach from Bulk Calculations}},\ }\href
  {https://doi.org/10.7566/JPSJ.82.023602} {\bibfield  {journal} {\bibinfo
  {journal} {J. Phys. Soc. Jpn.}\ }\textbf {\bibinfo {volume} {82}},\ \bibinfo
  {pages} {023602} (\bibinfo {year} {2013})}\BibitemShut {NoStop}%
\bibitem [{\citenamefont {{Ando}}\ and\ \citenamefont
  {{Fu}}(2015)}]{FuAndo2015}%
  \BibitemOpen
  \bibfield  {author} {\bibinfo {author} {\bibfnamefont {Y.}~\bibnamefont
  {{Ando}}}\ and\ \bibinfo {author} {\bibfnamefont {L.}~\bibnamefont {{Fu}}},\
  }\bibfield  {title} {\bibinfo {title} {{Topological Crystalline Insulators
  and Topological Superconductors: From Concepts to Materials}},\ }\href
  {https://doi.org/10.1146/annurev-conmatphys-031214-014501} {\bibfield
  {journal} {\bibinfo  {journal} {Annu. Rev. Condens. Matter Phys.}\ }\textbf
  {\bibinfo {volume} {6}},\ \bibinfo {pages} {361} (\bibinfo {year}
  {2015})}\BibitemShut {NoStop}%
\bibitem [{\citenamefont {Sato}\ and\ \citenamefont {Ando}(2017)}]{Sato2017}%
  \BibitemOpen
  \bibfield  {author} {\bibinfo {author} {\bibfnamefont {M.}~\bibnamefont
  {Sato}}\ and\ \bibinfo {author} {\bibfnamefont {Y.}~\bibnamefont {Ando}},\
  }\bibfield  {title} {\bibinfo {title} {{Topological superconductors: a
  review}},\ }\href {https://doi.org/10.1088/1361-6633/aa6ac7} {\bibfield
  {journal} {\bibinfo  {journal} {Rep. Prog. Phys.}\ }\textbf {\bibinfo
  {volume} {80}},\ \bibinfo {pages} {076501} (\bibinfo {year}
  {2017})}\BibitemShut {NoStop}%
\bibitem [{\citenamefont {Senthil}\ \emph {et~al.}(1999)\citenamefont
  {Senthil}, \citenamefont {Marston},\ and\ \citenamefont
  {Fisher}}]{senthilmarston1999}%
  \BibitemOpen
  \bibfield  {author} {\bibinfo {author} {\bibfnamefont {T.}~\bibnamefont
  {Senthil}}, \bibinfo {author} {\bibfnamefont {J.~B.}\ \bibnamefont
  {Marston}},\ and\ \bibinfo {author} {\bibfnamefont {M.~P.~A.}\ \bibnamefont
  {Fisher}},\ }\bibfield  {title} {\bibinfo {title} {{Spin quantum Hall effect
  in unconventional superconductors}},\ }\href
  {https://doi.org/10.1103/PhysRevB.60.4245} {\bibfield  {journal} {\bibinfo
  {journal} {Phys. Rev. B}\ }\textbf {\bibinfo {volume} {60}},\ \bibinfo
  {pages} {4245} (\bibinfo {year} {1999})}\BibitemShut {NoStop}%
\bibitem [{\citenamefont {Groth}\ \emph {et~al.}(2014)\citenamefont {Groth},
  \citenamefont {Wimmer}, \citenamefont {Akhmerov},\ and\ \citenamefont
  {Waintal}}]{Groth2014}%
  \BibitemOpen
  \bibfield  {author} {\bibinfo {author} {\bibfnamefont {C.~W.}\ \bibnamefont
  {Groth}}, \bibinfo {author} {\bibfnamefont {M.}~\bibnamefont {Wimmer}},
  \bibinfo {author} {\bibfnamefont {A.~R.}\ \bibnamefont {Akhmerov}},\ and\
  \bibinfo {author} {\bibfnamefont {X.}~\bibnamefont {Waintal}},\ }\bibfield
  {title} {\bibinfo {title} {{Kwant: a software package for quantum
  transport}},\ }\href {https://doi.org/10.1088/1367-2630/16/6/063065}
  {\bibfield  {journal} {\bibinfo  {journal} {New J. Phys.}\ }\textbf {\bibinfo
  {volume} {16}},\ \bibinfo {pages} {063065} (\bibinfo {year}
  {2014})}\BibitemShut {NoStop}%
\bibitem [{\citenamefont {Qi}\ \emph {et~al.}(2006)\citenamefont {Qi},
  \citenamefont {Wu},\ and\ \citenamefont {Zhang}}]{Qi2006}%
  \BibitemOpen
  \bibfield  {author} {\bibinfo {author} {\bibfnamefont {X.-L.}\ \bibnamefont
  {Qi}}, \bibinfo {author} {\bibfnamefont {Y.-S.}\ \bibnamefont {Wu}},\ and\
  \bibinfo {author} {\bibfnamefont {S.-C.}\ \bibnamefont {Zhang}},\ }\bibfield
  {title} {\bibinfo {title} {{Topological quantization of the spin Hall effect
  in two-dimensional paramagnetic semiconductors}},\ }\href
  {https://doi.org/10.1103/PhysRevB.74.085308} {\bibfield  {journal} {\bibinfo
  {journal} {Phys. Rev. B}\ }\textbf {\bibinfo {volume} {74}},\ \bibinfo
  {pages} {085308} (\bibinfo {year} {2006})}\BibitemShut {NoStop}%
\bibitem [{\citenamefont {Thouless}\ \emph {et~al.}(1982)\citenamefont
  {Thouless}, \citenamefont {Kohmoto}, \citenamefont {Nightingale},\ and\
  \citenamefont {den Nijs}}]{Thouless1982}%
  \BibitemOpen
  \bibfield  {author} {\bibinfo {author} {\bibfnamefont {D.~J.}\ \bibnamefont
  {Thouless}}, \bibinfo {author} {\bibfnamefont {M.}~\bibnamefont {Kohmoto}},
  \bibinfo {author} {\bibfnamefont {M.~P.}\ \bibnamefont {Nightingale}},\ and\
  \bibinfo {author} {\bibfnamefont {M.}~\bibnamefont {den Nijs}},\ }\bibfield
  {title} {\bibinfo {title} {{Quantized Hall Conductance in a Two-Dimensional
  Periodic Potential}},\ }\href {https://doi.org/10.1103/PhysRevLett.49.405}
  {\bibfield  {journal} {\bibinfo  {journal} {Phys. Rev. Lett.}\ }\textbf
  {\bibinfo {volume} {49}},\ \bibinfo {pages} {405} (\bibinfo {year}
  {1982})}\BibitemShut {NoStop}%
\bibitem [{\citenamefont {Long}\ \emph {et~al.}(2011)\citenamefont {Long},
  \citenamefont {Zhang},\ and\ \citenamefont {Sun}}]{Long2011}%
  \BibitemOpen
  \bibfield  {author} {\bibinfo {author} {\bibfnamefont {W.}~\bibnamefont
  {Long}}, \bibinfo {author} {\bibfnamefont {H.}~\bibnamefont {Zhang}},\ and\
  \bibinfo {author} {\bibfnamefont {Q.-f.}\ \bibnamefont {Sun}},\ }\bibfield
  {title} {\bibinfo {title} {{Quantum thermal Hall effect in graphene}},\
  }\href {https://doi.org/10.1103/PhysRevB.84.075416} {\bibfield  {journal}
  {\bibinfo  {journal} {Phys. Rev. B}\ }\textbf {\bibinfo {volume} {84}},\
  \bibinfo {pages} {075416} (\bibinfo {year} {2011})}\BibitemShut {NoStop}%
\bibitem [{\citenamefont {Fulga}\ \emph {et~al.}(2020)\citenamefont {Fulga},
  \citenamefont {Oreg}, \citenamefont {Mirlin}, \citenamefont {Stern},\ and\
  \citenamefont {Mross}}]{Fulga2020}%
  \BibitemOpen
  \bibfield  {author} {\bibinfo {author} {\bibfnamefont {I.~C.}\ \bibnamefont
  {Fulga}}, \bibinfo {author} {\bibfnamefont {Y.}~\bibnamefont {Oreg}},
  \bibinfo {author} {\bibfnamefont {A.~D.}\ \bibnamefont {Mirlin}}, \bibinfo
  {author} {\bibfnamefont {A.}~\bibnamefont {Stern}},\ and\ \bibinfo {author}
  {\bibfnamefont {D.~F.}\ \bibnamefont {Mross}},\ }\bibfield  {title} {\bibinfo
  {title} {{Temperature Enhancement of Thermal Hall Conductance
  Quantization}},\ }\href {https://doi.org/10.1103/PhysRevLett.125.236802}
  {\bibfield  {journal} {\bibinfo  {journal} {Phys. Rev. Lett.}\ }\textbf
  {\bibinfo {volume} {125}},\ \bibinfo {pages} {236802} (\bibinfo {year}
  {2020})}\BibitemShut {NoStop}%
\bibitem [{\citenamefont {Das}\ \emph {et~al.}(2023)\citenamefont {Das},
  \citenamefont {Manna},\ and\ \citenamefont {Roy}}]{Das2023}%
  \BibitemOpen
  \bibfield  {author} {\bibinfo {author} {\bibfnamefont {S.~K.}\ \bibnamefont
  {Das}}, \bibinfo {author} {\bibfnamefont {S.}~\bibnamefont {Manna}},\ and\
  \bibinfo {author} {\bibfnamefont {B.}~\bibnamefont {Roy}},\ }\bibfield
  {title} {\bibinfo {title} {{Topologically distinct atomic insulators}},\
  }\href {https://doi.org/10.1103/PhysRevB.108.L041301} {\bibfield  {journal}
  {\bibinfo  {journal} {Phys. Rev. B}\ }\textbf {\bibinfo {volume} {108}},\
  \bibinfo {pages} {L041301} (\bibinfo {year} {2023})}\BibitemShut {NoStop}%
\bibitem [{\citenamefont {Senthil}\ and\ \citenamefont
  {Fisher}(2000)}]{Senthil2000}%
  \BibitemOpen
  \bibfield  {author} {\bibinfo {author} {\bibfnamefont {T.}~\bibnamefont
  {Senthil}}\ and\ \bibinfo {author} {\bibfnamefont {M.~P.~A.}\ \bibnamefont
  {Fisher}},\ }\bibfield  {title} {\bibinfo {title} {{Quasiparticle
  localization in superconductors with spin-orbit scattering}},\ }\href
  {https://doi.org/10.1103/PhysRevB.61.9690} {\bibfield  {journal} {\bibinfo
  {journal} {Phys. Rev. B}\ }\textbf {\bibinfo {volume} {61}},\ \bibinfo
  {pages} {9690} (\bibinfo {year} {2000})}\BibitemShut {NoStop}%
\bibitem [{\citenamefont {Nomura}\ \emph {et~al.}(2007)\citenamefont {Nomura},
  \citenamefont {Koshino},\ and\ \citenamefont {Ryu}}]{Nomura2007}%
  \BibitemOpen
  \bibfield  {author} {\bibinfo {author} {\bibfnamefont {K.}~\bibnamefont
  {Nomura}}, \bibinfo {author} {\bibfnamefont {M.}~\bibnamefont {Koshino}},\
  and\ \bibinfo {author} {\bibfnamefont {S.}~\bibnamefont {Ryu}},\ }\bibfield
  {title} {\bibinfo {title} {{Topological Delocalization of Two-Dimensional
  Massless Dirac Fermions}},\ }\href
  {https://doi.org/10.1103/PhysRevLett.99.146806} {\bibfield  {journal}
  {\bibinfo  {journal} {Phys. Rev. Lett.}\ }\textbf {\bibinfo {volume} {99}},\
  \bibinfo {pages} {146806} (\bibinfo {year} {2007})}\BibitemShut {NoStop}%
\bibitem [{\citenamefont {Shindou}\ and\ \citenamefont
  {Murakami}(2009)}]{Shindou2009}%
  \BibitemOpen
  \bibfield  {author} {\bibinfo {author} {\bibfnamefont {R.}~\bibnamefont
  {Shindou}}\ and\ \bibinfo {author} {\bibfnamefont {S.}~\bibnamefont
  {Murakami}},\ }\bibfield  {title} {\bibinfo {title} {{Effects of disorder in
  three-dimensional ${Z}_{2}$ quantum spin {Hall} systems}},\ }\href
  {https://doi.org/10.1103/PhysRevB.79.045321} {\bibfield  {journal} {\bibinfo
  {journal} {Phys. Rev. B}\ }\textbf {\bibinfo {volume} {79}},\ \bibinfo
  {pages} {045321} (\bibinfo {year} {2009})}\BibitemShut {NoStop}%
\bibitem [{\citenamefont {Ostrovsky}\ \emph {et~al.}(2010)\citenamefont
  {Ostrovsky}, \citenamefont {Gornyi},\ and\ \citenamefont
  {Mirlin}}]{Mirlin2010}%
  \BibitemOpen
  \bibfield  {author} {\bibinfo {author} {\bibfnamefont {P.~M.}\ \bibnamefont
  {Ostrovsky}}, \bibinfo {author} {\bibfnamefont {I.~V.}\ \bibnamefont
  {Gornyi}},\ and\ \bibinfo {author} {\bibfnamefont {A.~D.}\ \bibnamefont
  {Mirlin}},\ }\bibfield  {title} {\bibinfo {title} {{Interaction-Induced
  Criticality in ${\mathbb{Z}}_{2}$ Topological Insulators}},\ }\href
  {https://doi.org/10.1103/PhysRevLett.105.036803} {\bibfield  {journal}
  {\bibinfo  {journal} {Phys. Rev. Lett.}\ }\textbf {\bibinfo {volume} {105}},\
  \bibinfo {pages} {036803} (\bibinfo {year} {2010})}\BibitemShut {NoStop}%
\bibitem [{\citenamefont {Goswami}\ and\ \citenamefont
  {Chakravarty}(2011)}]{Goswami2011}%
  \BibitemOpen
  \bibfield  {author} {\bibinfo {author} {\bibfnamefont {P.}~\bibnamefont
  {Goswami}}\ and\ \bibinfo {author} {\bibfnamefont {S.}~\bibnamefont
  {Chakravarty}},\ }\bibfield  {title} {\bibinfo {title} {{Quantum Criticality
  between Topological and Band Insulators in $3+1$ Dimensions}},\ }\href
  {https://doi.org/10.1103/PhysRevLett.107.196803} {\bibfield  {journal}
  {\bibinfo  {journal} {Phys. Rev. Lett.}\ }\textbf {\bibinfo {volume} {107}},\
  \bibinfo {pages} {196803} (\bibinfo {year} {2011})}\BibitemShut {NoStop}%
\bibitem [{\citenamefont {Ringel}\ \emph {et~al.}(2012)\citenamefont {Ringel},
  \citenamefont {Kraus},\ and\ \citenamefont {Stern}}]{Ringel2012}%
  \BibitemOpen
  \bibfield  {author} {\bibinfo {author} {\bibfnamefont {Z.}~\bibnamefont
  {Ringel}}, \bibinfo {author} {\bibfnamefont {Y.~E.}\ \bibnamefont {Kraus}},\
  and\ \bibinfo {author} {\bibfnamefont {A.}~\bibnamefont {Stern}},\ }\bibfield
   {title} {\bibinfo {title} {{Strong side of weak topological insulators}},\
  }\href {https://doi.org/10.1103/PhysRevB.86.045102} {\bibfield  {journal}
  {\bibinfo  {journal} {Phys. Rev. B}\ }\textbf {\bibinfo {volume} {86}},\
  \bibinfo {pages} {045102} (\bibinfo {year} {2012})}\BibitemShut {NoStop}%
\bibitem [{\citenamefont {Kobayashi}\ \emph {et~al.}(2013)\citenamefont
  {Kobayashi}, \citenamefont {Ohtsuki},\ and\ \citenamefont
  {Imura}}]{Kobayashi2013}%
  \BibitemOpen
  \bibfield  {author} {\bibinfo {author} {\bibfnamefont {K.}~\bibnamefont
  {Kobayashi}}, \bibinfo {author} {\bibfnamefont {T.}~\bibnamefont {Ohtsuki}},\
  and\ \bibinfo {author} {\bibfnamefont {K.-I.}\ \bibnamefont {Imura}},\
  }\bibfield  {title} {\bibinfo {title} {{Disordered Weak and Strong
  Topological Insulators}},\ }\href
  {https://doi.org/10.1103/PhysRevLett.110.236803} {\bibfield  {journal}
  {\bibinfo  {journal} {Phys. Rev. Lett.}\ }\textbf {\bibinfo {volume} {110}},\
  \bibinfo {pages} {236803} (\bibinfo {year} {2013})}\BibitemShut {NoStop}%
\bibitem [{\citenamefont {Fulga}\ \emph {et~al.}(2014)\citenamefont {Fulga},
  \citenamefont {van Heck}, \citenamefont {Edge},\ and\ \citenamefont
  {Akhmerov}}]{Fulga2014}%
  \BibitemOpen
  \bibfield  {author} {\bibinfo {author} {\bibfnamefont {I.~C.}\ \bibnamefont
  {Fulga}}, \bibinfo {author} {\bibfnamefont {B.}~\bibnamefont {van Heck}},
  \bibinfo {author} {\bibfnamefont {J.~M.}\ \bibnamefont {Edge}},\ and\
  \bibinfo {author} {\bibfnamefont {A.~R.}\ \bibnamefont {Akhmerov}},\
  }\bibfield  {title} {\bibinfo {title} {{Statistical topological
  insulators}},\ }\href {https://doi.org/10.1103/PhysRevB.89.155424} {\bibfield
   {journal} {\bibinfo  {journal} {Phys. Rev. B}\ }\textbf {\bibinfo {volume}
  {89}},\ \bibinfo {pages} {155424} (\bibinfo {year} {2014})}\BibitemShut
  {NoStop}%
\bibitem [{\citenamefont {Morimoto}\ \emph {et~al.}(2015)\citenamefont
  {Morimoto}, \citenamefont {Furusaki},\ and\ \citenamefont
  {Mudry}}]{ChrisMudry2015}%
  \BibitemOpen
  \bibfield  {author} {\bibinfo {author} {\bibfnamefont {T.}~\bibnamefont
  {Morimoto}}, \bibinfo {author} {\bibfnamefont {A.}~\bibnamefont {Furusaki}},\
  and\ \bibinfo {author} {\bibfnamefont {C.}~\bibnamefont {Mudry}},\ }\bibfield
   {title} {\bibinfo {title} {{Anderson localization and the topology of
  classifying spaces}},\ }\href {https://doi.org/10.1103/PhysRevB.91.235111}
  {\bibfield  {journal} {\bibinfo  {journal} {Phys. Rev. B}\ }\textbf {\bibinfo
  {volume} {91}},\ \bibinfo {pages} {235111} (\bibinfo {year}
  {2015})}\BibitemShut {NoStop}%
\bibitem [{\citenamefont {Roy}\ \emph {et~al.}(2017)\citenamefont {Roy},
  \citenamefont {Alavirad},\ and\ \citenamefont {Sau}}]{Alavirad2017}%
  \BibitemOpen
  \bibfield  {author} {\bibinfo {author} {\bibfnamefont {B.}~\bibnamefont
  {Roy}}, \bibinfo {author} {\bibfnamefont {Y.}~\bibnamefont {Alavirad}},\ and\
  \bibinfo {author} {\bibfnamefont {J.~D.}\ \bibnamefont {Sau}},\ }\bibfield
  {title} {\bibinfo {title} {{Global Phase Diagram of a Three-Dimensional Dirty
  Topological Superconductor}},\ }\href
  {https://doi.org/10.1103/PhysRevLett.118.227002} {\bibfield  {journal}
  {\bibinfo  {journal} {Phys. Rev. Lett.}\ }\textbf {\bibinfo {volume} {118}},\
  \bibinfo {pages} {227002} (\bibinfo {year} {2017})}\BibitemShut {NoStop}%
\bibitem [{\citenamefont {Roy}\ \emph {et~al.}(2019)\citenamefont {Roy},
  \citenamefont {Ghorashi}, \citenamefont {Foster},\ and\ \citenamefont
  {Nevidomskyy}}]{Roy2019}%
  \BibitemOpen
  \bibfield  {author} {\bibinfo {author} {\bibfnamefont {B.}~\bibnamefont
  {Roy}}, \bibinfo {author} {\bibfnamefont {S.~A.~A.}\ \bibnamefont
  {Ghorashi}}, \bibinfo {author} {\bibfnamefont {M.~S.}\ \bibnamefont
  {Foster}},\ and\ \bibinfo {author} {\bibfnamefont {A.~H.}\ \bibnamefont
  {Nevidomskyy}},\ }\bibfield  {title} {\bibinfo {title} {{Topological
  superconductivity of spin-$3/2$ carriers in a three-dimensional doped
  {Luttinger} semimetal}},\ }\href {https://doi.org/10.1103/PhysRevB.99.054505}
  {\bibfield  {journal} {\bibinfo  {journal} {Phys. Rev. B}\ }\textbf {\bibinfo
  {volume} {99}},\ \bibinfo {pages} {054505} (\bibinfo {year}
  {2019})}\BibitemShut {NoStop}%
\bibitem [{\citenamefont {{Banerjee}}\ \emph {et~al.}(2017)\citenamefont
  {{Banerjee}}, \citenamefont {{Heiblum}}, \citenamefont {{Rosenblatt}},
  \citenamefont {{Oreg}}, \citenamefont {{Feldman}}, \citenamefont {{Stern}},\
  and\ \citenamefont {{Umansky}}}]{Banerjee2018a}%
  \BibitemOpen
  \bibfield  {author} {\bibinfo {author} {\bibfnamefont {M.}~\bibnamefont
  {{Banerjee}}}, \bibinfo {author} {\bibfnamefont {M.}~\bibnamefont
  {{Heiblum}}}, \bibinfo {author} {\bibfnamefont {A.}~\bibnamefont
  {{Rosenblatt}}}, \bibinfo {author} {\bibfnamefont {Y.}~\bibnamefont
  {{Oreg}}}, \bibinfo {author} {\bibfnamefont {D.~E.}\ \bibnamefont
  {{Feldman}}}, \bibinfo {author} {\bibfnamefont {A.}~\bibnamefont {{Stern}}},\
  and\ \bibinfo {author} {\bibfnamefont {V.}~\bibnamefont {{Umansky}}},\
  }\bibfield  {title} {\bibinfo {title} {{Observed quantization of anyonic heat
  flow}},\ }\href {https://doi.org/10.1038/nature22052} {\bibfield  {journal}
  {\bibinfo  {journal} {\nat}\ }\textbf {\bibinfo {volume} {545}},\ \bibinfo
  {pages} {75} (\bibinfo {year} {2017})}\BibitemShut {NoStop}%
\bibitem [{\citenamefont {{Banerjee}}\ \emph {et~al.}(2018)\citenamefont
  {{Banerjee}}, \citenamefont {{Heiblum}}, \citenamefont {{Umansky}},
  \citenamefont {{Feldman}}, \citenamefont {{Oreg}},\ and\ \citenamefont
  {{Stern}}}]{Banerjee2018b}%
  \BibitemOpen
  \bibfield  {author} {\bibinfo {author} {\bibfnamefont {M.}~\bibnamefont
  {{Banerjee}}}, \bibinfo {author} {\bibfnamefont {M.}~\bibnamefont
  {{Heiblum}}}, \bibinfo {author} {\bibfnamefont {V.}~\bibnamefont
  {{Umansky}}}, \bibinfo {author} {\bibfnamefont {D.~E.}\ \bibnamefont
  {{Feldman}}}, \bibinfo {author} {\bibfnamefont {Y.}~\bibnamefont {{Oreg}}},\
  and\ \bibinfo {author} {\bibfnamefont {A.}~\bibnamefont {{Stern}}},\
  }\bibfield  {title} {\bibinfo {title} {{Observation of half-integer thermal
  Hall conductance}},\ }\href {https://doi.org/10.1038/s41586-018-0184-1}
  {\bibfield  {journal} {\bibinfo  {journal} {\nat}\ }\textbf {\bibinfo
  {volume} {559}},\ \bibinfo {pages} {205} (\bibinfo {year}
  {2018})}\BibitemShut {NoStop}%
\bibitem [{\citenamefont {Kasahara}\ \emph {et~al.}(2018)\citenamefont
  {Kasahara}, \citenamefont {Ohnishi}, \citenamefont {Mizukami}, \citenamefont
  {Tanaka}, \citenamefont {Ma}, \citenamefont {Sugii}, \citenamefont {Kurita},
  \citenamefont {Tanaka}, \citenamefont {Nasu}, \citenamefont {Motome},
  \citenamefont {Shibauchi},\ and\ \citenamefont {Matsuda}}]{Kasahara2018}%
  \BibitemOpen
  \bibfield  {author} {\bibinfo {author} {\bibfnamefont {Y.}~\bibnamefont
  {Kasahara}}, \bibinfo {author} {\bibfnamefont {T.}~\bibnamefont {Ohnishi}},
  \bibinfo {author} {\bibfnamefont {Y.}~\bibnamefont {Mizukami}}, \bibinfo
  {author} {\bibfnamefont {O.}~\bibnamefont {Tanaka}}, \bibinfo {author}
  {\bibfnamefont {S.}~\bibnamefont {Ma}}, \bibinfo {author} {\bibfnamefont
  {K.}~\bibnamefont {Sugii}}, \bibinfo {author} {\bibfnamefont
  {N.}~\bibnamefont {Kurita}}, \bibinfo {author} {\bibfnamefont
  {H.}~\bibnamefont {Tanaka}}, \bibinfo {author} {\bibfnamefont
  {J.}~\bibnamefont {Nasu}}, \bibinfo {author} {\bibfnamefont {Y.}~\bibnamefont
  {Motome}}, \bibinfo {author} {\bibfnamefont {T.}~\bibnamefont {Shibauchi}},\
  and\ \bibinfo {author} {\bibfnamefont {Y.}~\bibnamefont {Matsuda}},\
  }\bibfield  {title} {\bibinfo {title} {{Majorana quantization and
  half-integer thermal quantum Hall effect in a Kitaev spin liquid}},\ }\href
  {https://doi.org/10.1038/s41586-018-0274-0} {\bibfield  {journal} {\bibinfo
  {journal} {\nat}\ }\textbf {\bibinfo {volume} {559}},\ \bibinfo {pages} {227}
  (\bibinfo {year} {2018})}\BibitemShut {NoStop}%
\bibitem [{\citenamefont {{Srivastav}}\ \emph {et~al.}(2019)\citenamefont
  {{Srivastav}}, \citenamefont {{Sahu}}, \citenamefont {{Watanabe}},
  \citenamefont {{Taniguchi}}, \citenamefont {{Banerjee}},\ and\ \citenamefont
  {{Das}}}]{Srivastav2019}%
  \BibitemOpen
  \bibfield  {author} {\bibinfo {author} {\bibfnamefont {S.~K.}\ \bibnamefont
  {{Srivastav}}}, \bibinfo {author} {\bibfnamefont {M.~R.}\ \bibnamefont
  {{Sahu}}}, \bibinfo {author} {\bibfnamefont {K.}~\bibnamefont {{Watanabe}}},
  \bibinfo {author} {\bibfnamefont {T.}~\bibnamefont {{Taniguchi}}}, \bibinfo
  {author} {\bibfnamefont {S.}~\bibnamefont {{Banerjee}}},\ and\ \bibinfo
  {author} {\bibfnamefont {A.}~\bibnamefont {{Das}}},\ }\bibfield  {title}
  {\bibinfo {title} {{Universal quantized thermal conductance in graphene}},\
  }\href {https://doi.org/10.1126/sciadv.aaw5798} {\bibfield  {journal}
  {\bibinfo  {journal} {Sci. Adv.}\ }\textbf {\bibinfo {volume} {5}},\ \bibinfo
  {eid} {eaaw5798} (\bibinfo {year} {2019})}\BibitemShut {NoStop}%
\bibitem [{\citenamefont {Le~Breton}\ \emph {et~al.}(2022)\citenamefont
  {Le~Breton}, \citenamefont {Delagrange}, \citenamefont {Hong}, \citenamefont
  {Garg}, \citenamefont {Watanabe}, \citenamefont {Taniguchi}, \citenamefont
  {Ribeiro-Palau}, \citenamefont {Roulleau}, \citenamefont {Roche},\ and\
  \citenamefont {Parmentier}}]{Breton2022}%
  \BibitemOpen
  \bibfield  {author} {\bibinfo {author} {\bibfnamefont {G.}~\bibnamefont
  {Le~Breton}}, \bibinfo {author} {\bibfnamefont {R.}~\bibnamefont
  {Delagrange}}, \bibinfo {author} {\bibfnamefont {Y.}~\bibnamefont {Hong}},
  \bibinfo {author} {\bibfnamefont {M.}~\bibnamefont {Garg}}, \bibinfo {author}
  {\bibfnamefont {K.}~\bibnamefont {Watanabe}}, \bibinfo {author}
  {\bibfnamefont {T.}~\bibnamefont {Taniguchi}}, \bibinfo {author}
  {\bibfnamefont {R.}~\bibnamefont {Ribeiro-Palau}}, \bibinfo {author}
  {\bibfnamefont {P.}~\bibnamefont {Roulleau}}, \bibinfo {author}
  {\bibfnamefont {P.}~\bibnamefont {Roche}},\ and\ \bibinfo {author}
  {\bibfnamefont {F.~D.}\ \bibnamefont {Parmentier}},\ }\bibfield  {title}
  {\bibinfo {title} {{Heat Equilibration of Integer and Fractional Quantum Hall
  Edge Modes in Graphene}},\ }\href
  {https://doi.org/10.1103/PhysRevLett.129.116803} {\bibfield  {journal}
  {\bibinfo  {journal} {Phys. Rev. Lett.}\ }\textbf {\bibinfo {volume} {129}},\
  \bibinfo {pages} {116803} (\bibinfo {year} {2022})}\BibitemShut {NoStop}%
\bibitem [{\citenamefont {{S. K. Das}}\ and\ \citenamefont {{B.
  Roy}}(2023)}]{zenodoTSCSanjib}%
  \BibitemOpen
  \bibfield  {author} {\bibinfo {author} {\bibnamefont {{S. K. Das}}}\ and\
  \bibinfo {author} {\bibnamefont {{B. Roy}}},\ }\bibfield  {title} {\bibinfo
  {title} {{Quantized thermal and spin transports of dirty planar topological
  superconductors}}\ }\href {https://doi.org/10.5281/zenodo.8370723}
  {10.5281/zenodo.8370723} (\bibinfo {year} {2023})\BibitemShut {NoStop}%
\bibitem [{\citenamefont {Manna}\ \emph {et~al.}()\citenamefont {Manna},
  \citenamefont {Das},\ and\ \citenamefont {Roy}}]{MannaDas2022}%
  \BibitemOpen
  \bibfield  {author} {\bibinfo {author} {\bibfnamefont {S.}~\bibnamefont
  {Manna}}, \bibinfo {author} {\bibfnamefont {S.~K.}\ \bibnamefont {Das}},\
  and\ \bibinfo {author} {\bibfnamefont {B.}~\bibnamefont {Roy}},\ }\href@noop
  {} {\bibinfo {title} {{Noncrystalline topological superconductors}}},\
  \Eprint {https://arxiv.org/abs/arXiv:2207.02203} {arXiv:2207.02203}
  \BibitemShut {NoStop}%
\bibitem [{\citenamefont {Bianco}\ and\ \citenamefont
  {Resta}(2011)}]{Resta2011}%
  \BibitemOpen
  \bibfield  {author} {\bibinfo {author} {\bibfnamefont {R.}~\bibnamefont
  {Bianco}}\ and\ \bibinfo {author} {\bibfnamefont {R.}~\bibnamefont {Resta}},\
  }\bibfield  {title} {\bibinfo {title} {{Mapping topological order in
  coordinate space}},\ }\href {https://doi.org/10.1103/PhysRevB.84.241106}
  {\bibfield  {journal} {\bibinfo  {journal} {Phys. Rev. B}\ }\textbf {\bibinfo
  {volume} {84}},\ \bibinfo {pages} {241106} (\bibinfo {year}
  {2011})}\BibitemShut {NoStop}%
\bibitem [{\citenamefont {Chen}(2023)}]{WeiChen2023}%
  \BibitemOpen
  \bibfield  {author} {\bibinfo {author} {\bibfnamefont {W.}~\bibnamefont
  {Chen}},\ }\bibfield  {title} {\bibinfo {title} {{Universal topological
  marker}},\ }\href {https://doi.org/10.1103/PhysRevB.107.045111} {\bibfield
  {journal} {\bibinfo  {journal} {Phys. Rev. B}\ }\textbf {\bibinfo {volume}
  {107}},\ \bibinfo {pages} {045111} (\bibinfo {year} {2023})}\BibitemShut
  {NoStop}%
\bibitem [{\citenamefont {Mandal}\ \emph {et~al.}()\citenamefont {Mandal},
  \citenamefont {Drucker}, \citenamefont {Siriviboon}, \citenamefont {Nguyen},
  \citenamefont {Boonkird}, \citenamefont {Lamichhane}, \citenamefont {Okabe},
  \citenamefont {Chotrattanapituk},\ and\ \citenamefont {Li}}]{Mandal2023}%
  \BibitemOpen
  \bibfield  {author} {\bibinfo {author} {\bibfnamefont {M.}~\bibnamefont
  {Mandal}}, \bibinfo {author} {\bibfnamefont {N.~C.}\ \bibnamefont {Drucker}},
  \bibinfo {author} {\bibfnamefont {P.}~\bibnamefont {Siriviboon}}, \bibinfo
  {author} {\bibfnamefont {T.}~\bibnamefont {Nguyen}}, \bibinfo {author}
  {\bibfnamefont {T.}~\bibnamefont {Boonkird}}, \bibinfo {author}
  {\bibfnamefont {T.~N.}\ \bibnamefont {Lamichhane}}, \bibinfo {author}
  {\bibfnamefont {R.}~\bibnamefont {Okabe}}, \bibinfo {author} {\bibfnamefont
  {A.}~\bibnamefont {Chotrattanapituk}},\ and\ \bibinfo {author} {\bibfnamefont
  {M.}~\bibnamefont {Li}},\ }\href@noop {} {\bibinfo {title} {{Topological
  superconductors from a materials perspective}}},\ \Eprint
  {https://arxiv.org/abs/arXiv:2303.15581} {arXiv:2303.15581} \BibitemShut
  {NoStop}%
\bibitem [{\citenamefont {Mackenzie}\ and\ \citenamefont
  {Maeno}(2003)}]{MackenzieMaeno2003}%
  \BibitemOpen
  \bibfield  {author} {\bibinfo {author} {\bibfnamefont {A.~P.}\ \bibnamefont
  {Mackenzie}}\ and\ \bibinfo {author} {\bibfnamefont {Y.}~\bibnamefont
  {Maeno}},\ }\bibfield  {title} {\bibinfo {title} {{The superconductivity of
  ${\mathrm{Sr}}_{2}{\mathrm{RuO}}_{4}$ and the physics of spin-triplet
  pairing}},\ }\href {https://doi.org/10.1103/RevModPhys.75.657} {\bibfield
  {journal} {\bibinfo  {journal} {Rev. Mod. Phys.}\ }\textbf {\bibinfo {volume}
  {75}},\ \bibinfo {pages} {657} (\bibinfo {year} {2003})}\BibitemShut
  {NoStop}%
\bibitem [{\citenamefont {Kallin}(2012)}]{Kallin2012}%
  \BibitemOpen
  \bibfield  {author} {\bibinfo {author} {\bibfnamefont {C.}~\bibnamefont
  {Kallin}},\ }\bibfield  {title} {\bibinfo {title} {{Chiral p-wave order in
  {Sr$_2$RuO$_4$}}},\ }\href {https://doi.org/10.1088/0034-4885/75/4/042501}
  {\bibfield  {journal} {\bibinfo  {journal} {Rep. Prog. Phys.}\ }\textbf
  {\bibinfo {volume} {75}},\ \bibinfo {pages} {042501} (\bibinfo {year}
  {2012})}\BibitemShut {NoStop}%
\bibitem [{\citenamefont {{Maeno}}\ \emph {et~al.}(2012)\citenamefont
  {{Maeno}}, \citenamefont {{Kittaka}}, \citenamefont {{Nomura}}, \citenamefont
  {{Yonezawa}},\ and\ \citenamefont {{Ishida}}}]{Maenoetal2012}%
  \BibitemOpen
  \bibfield  {author} {\bibinfo {author} {\bibfnamefont {Y.}~\bibnamefont
  {{Maeno}}}, \bibinfo {author} {\bibfnamefont {S.}~\bibnamefont {{Kittaka}}},
  \bibinfo {author} {\bibfnamefont {T.}~\bibnamefont {{Nomura}}}, \bibinfo
  {author} {\bibfnamefont {S.}~\bibnamefont {{Yonezawa}}},\ and\ \bibinfo
  {author} {\bibfnamefont {K.}~\bibnamefont {{Ishida}}},\ }\bibfield  {title}
  {\bibinfo {title} {{Evaluation of Spin-Triplet Superconductivity in
  Sr$_{2}$RuO$_{4}$}},\ }\href {https://doi.org/10.1143/JPSJ.81.011009}
  {\bibfield  {journal} {\bibinfo  {journal} {J. Phys. Soc. Jpn.}\ }\textbf
  {\bibinfo {volume} {81}},\ \bibinfo {pages} {011009} (\bibinfo {year}
  {2012})}\BibitemShut {NoStop}%
\bibitem [{\citenamefont {Fu}\ and\ \citenamefont {Berg}(2010)}]{FuBerg2010}%
  \BibitemOpen
  \bibfield  {author} {\bibinfo {author} {\bibfnamefont {L.}~\bibnamefont
  {Fu}}\ and\ \bibinfo {author} {\bibfnamefont {E.}~\bibnamefont {Berg}},\
  }\bibfield  {title} {\bibinfo {title} {{Odd-Parity Topological
  Superconductors: Theory and Application to
  ${\mathrm{Cu}}_{x}{\mathrm{Bi}}_{2}{\mathrm{Se}}_{3}$}},\ }\href
  {https://doi.org/10.1103/PhysRevLett.105.097001} {\bibfield  {journal}
  {\bibinfo  {journal} {Phys. Rev. Lett.}\ }\textbf {\bibinfo {volume} {105}},\
  \bibinfo {pages} {097001} (\bibinfo {year} {2010})}\BibitemShut {NoStop}%
\bibitem [{\citenamefont {Fu}(2014)}]{LiangFu2014}%
  \BibitemOpen
  \bibfield  {author} {\bibinfo {author} {\bibfnamefont {L.}~\bibnamefont
  {Fu}},\ }\bibfield  {title} {\bibinfo {title} {{Odd-parity topological
  superconductor with nematic order: Application to
  ${\mathrm{Cu}}_{x}{\mathrm{Bi}}_{2}{\mathrm{Se}}_{3}$}},\ }\href
  {https://doi.org/10.1103/PhysRevB.90.100509} {\bibfield  {journal} {\bibinfo
  {journal} {Phys. Rev. B}\ }\textbf {\bibinfo {volume} {90}},\ \bibinfo
  {pages} {100509} (\bibinfo {year} {2014})}\BibitemShut {NoStop}%
\bibitem [{\citenamefont {Roy}(2020)}]{BRoy2020}%
  \BibitemOpen
  \bibfield  {author} {\bibinfo {author} {\bibfnamefont {B.}~\bibnamefont
  {Roy}},\ }\bibfield  {title} {\bibinfo {title} {{Higher-order topological
  superconductors in $\mathcal{P}$-, $\mathcal{T}$-odd quadrupolar Dirac
  materials}},\ }\href {https://doi.org/10.1103/PhysRevB.101.220506} {\bibfield
   {journal} {\bibinfo  {journal} {Phys. Rev. B}\ }\textbf {\bibinfo {volume}
  {101}},\ \bibinfo {pages} {220506} (\bibinfo {year} {2020})}\BibitemShut
  {NoStop}%
\bibitem [{\citenamefont {Roy}\ and\ \citenamefont {Juri\ifmmode \check{c}\else
  \v{c}\fi{}i\ifmmode~\acute{c}\else \'{c}\fi{}}(2021)}]{RoyJuricic2021}%
  \BibitemOpen
  \bibfield  {author} {\bibinfo {author} {\bibfnamefont {B.}~\bibnamefont
  {Roy}}\ and\ \bibinfo {author} {\bibfnamefont {V.}~\bibnamefont {Juri\ifmmode
  \check{c}\else \v{c}\fi{}i\ifmmode~\acute{c}\else \'{c}\fi{}}},\ }\bibfield
  {title} {\bibinfo {title} {{Mixed-parity octupolar pairing and corner
  Majorana modes in three dimensions}},\ }\href
  {https://doi.org/10.1103/PhysRevB.104.L180503} {\bibfield  {journal}
  {\bibinfo  {journal} {Phys. Rev. B}\ }\textbf {\bibinfo {volume} {104}},\
  \bibinfo {pages} {L180503} (\bibinfo {year} {2021})}\BibitemShut {NoStop}%
\bibitem [{\citenamefont {Mandal}\ and\ \citenamefont
  {Roy}(2023)}]{MandalRoy2023}%
  \BibitemOpen
  \bibfield  {author} {\bibinfo {author} {\bibfnamefont {S.}~\bibnamefont
  {Mandal}}\ and\ \bibinfo {author} {\bibfnamefont {B.}~\bibnamefont {Roy}},\
  }\bibfield  {title} {\bibinfo {title} {{Polar hairs of mixed-parity nodal
  superconductors in Rarita-Schwinger-Weyl metals}},\ }\href
  {https://doi.org/10.1103/PhysRevB.107.L180502} {\bibfield  {journal}
  {\bibinfo  {journal} {Phys. Rev. B}\ }\textbf {\bibinfo {volume} {107}},\
  \bibinfo {pages} {L180502} (\bibinfo {year} {2023})}\BibitemShut {NoStop}%
\bibitem [{\citenamefont {Kriener}\ \emph {et~al.}(2011)\citenamefont
  {Kriener}, \citenamefont {Segawa}, \citenamefont {Ren}, \citenamefont
  {Sasaki},\ and\ \citenamefont {Ando}}]{Ando2011PRL}%
  \BibitemOpen
  \bibfield  {author} {\bibinfo {author} {\bibfnamefont {M.}~\bibnamefont
  {Kriener}}, \bibinfo {author} {\bibfnamefont {K.}~\bibnamefont {Segawa}},
  \bibinfo {author} {\bibfnamefont {Z.}~\bibnamefont {Ren}}, \bibinfo {author}
  {\bibfnamefont {S.}~\bibnamefont {Sasaki}},\ and\ \bibinfo {author}
  {\bibfnamefont {Y.}~\bibnamefont {Ando}},\ }\bibfield  {title} {\bibinfo
  {title} {{Bulk Superconducting Phase with a Full Energy Gap in the Doped
  Topological Insulator
  ${\mathrm{Cu}}_{x}{\mathrm{Bi}}_{2}{\mathrm{Se}}_{3}$}},\ }\href
  {https://doi.org/10.1103/PhysRevLett.106.127004} {\bibfield  {journal}
  {\bibinfo  {journal} {Phys. Rev. Lett.}\ }\textbf {\bibinfo {volume} {106}},\
  \bibinfo {pages} {127004} (\bibinfo {year} {2011})}\BibitemShut {NoStop}%
\bibitem [{\citenamefont {Sasaki}\ \emph {et~al.}(2012)\citenamefont {Sasaki},
  \citenamefont {Ren}, \citenamefont {Taskin}, \citenamefont {Segawa},
  \citenamefont {Fu},\ and\ \citenamefont {Ando}}]{Ando2012PRL}%
  \BibitemOpen
  \bibfield  {author} {\bibinfo {author} {\bibfnamefont {S.}~\bibnamefont
  {Sasaki}}, \bibinfo {author} {\bibfnamefont {Z.}~\bibnamefont {Ren}},
  \bibinfo {author} {\bibfnamefont {A.~A.}\ \bibnamefont {Taskin}}, \bibinfo
  {author} {\bibfnamefont {K.}~\bibnamefont {Segawa}}, \bibinfo {author}
  {\bibfnamefont {L.}~\bibnamefont {Fu}},\ and\ \bibinfo {author}
  {\bibfnamefont {Y.}~\bibnamefont {Ando}},\ }\bibfield  {title} {\bibinfo
  {title} {{Odd-Parity Pairing and Topological Superconductivity in a Strongly
  Spin-Orbit Coupled Semiconductor}},\ }\href
  {https://doi.org/10.1103/PhysRevLett.109.217004} {\bibfield  {journal}
  {\bibinfo  {journal} {Phys. Rev. Lett.}\ }\textbf {\bibinfo {volume} {109}},\
  \bibinfo {pages} {217004} (\bibinfo {year} {2012})}\BibitemShut {NoStop}%
\bibitem [{\citenamefont {Novak}\ \emph {et~al.}(2013)\citenamefont {Novak},
  \citenamefont {Sasaki}, \citenamefont {Kriener}, \citenamefont {Segawa},\
  and\ \citenamefont {Ando}}]{Ando2013PRB}%
  \BibitemOpen
  \bibfield  {author} {\bibinfo {author} {\bibfnamefont {M.}~\bibnamefont
  {Novak}}, \bibinfo {author} {\bibfnamefont {S.}~\bibnamefont {Sasaki}},
  \bibinfo {author} {\bibfnamefont {M.}~\bibnamefont {Kriener}}, \bibinfo
  {author} {\bibfnamefont {K.}~\bibnamefont {Segawa}},\ and\ \bibinfo {author}
  {\bibfnamefont {Y.}~\bibnamefont {Ando}},\ }\bibfield  {title} {\bibinfo
  {title} {{Unusual nature of fully gapped superconductivity in In-doped
  SnTe}},\ }\href {https://doi.org/10.1103/PhysRevB.88.140502} {\bibfield
  {journal} {\bibinfo  {journal} {Phys. Rev. B}\ }\textbf {\bibinfo {volume}
  {88}},\ \bibinfo {pages} {140502} (\bibinfo {year} {2013})}\BibitemShut
  {NoStop}%
\bibitem [{\citenamefont {{Yu}}\ \emph {et~al.}(2010)\citenamefont {{Yu}},
  \citenamefont {{Zhang}}, \citenamefont {{Zhang}}, \citenamefont {{Zhang}},
  \citenamefont {{Dai}},\ and\ \citenamefont {{Fang}}}]{ZhongFang2010}%
  \BibitemOpen
  \bibfield  {author} {\bibinfo {author} {\bibfnamefont {R.}~\bibnamefont
  {{Yu}}}, \bibinfo {author} {\bibfnamefont {W.}~\bibnamefont {{Zhang}}},
  \bibinfo {author} {\bibfnamefont {H.-J.}\ \bibnamefont {{Zhang}}}, \bibinfo
  {author} {\bibfnamefont {S.-C.}\ \bibnamefont {{Zhang}}}, \bibinfo {author}
  {\bibfnamefont {X.}~\bibnamefont {{Dai}}},\ and\ \bibinfo {author}
  {\bibfnamefont {Z.}~\bibnamefont {{Fang}}},\ }\bibfield  {title} {\bibinfo
  {title} {{Quantized Anomalous Hall Effect in Magnetic Topological
  Insulators}},\ }\href {https://doi.org/10.1126/science.1187485} {\bibfield
  {journal} {\bibinfo  {journal} {Science}\ }\textbf {\bibinfo {volume}
  {329}},\ \bibinfo {pages} {61} (\bibinfo {year} {2010})}\BibitemShut
  {NoStop}%
\bibitem [{\citenamefont {{Chang}}\ \emph {et~al.}(2013)\citenamefont
  {{Chang}}, \citenamefont {{Zhang}}, \citenamefont {{Feng}}, \citenamefont
  {{Shen}}, \citenamefont {{Zhang}}, \citenamefont {{Guo}}, \citenamefont
  {{Li}}, \citenamefont {{Ou}}, \citenamefont {{Wei}}, \citenamefont {{Wang}},
  \citenamefont {{Ji}}, \citenamefont {{Feng}}, \citenamefont {{Ji}},
  \citenamefont {{Chen}}, \citenamefont {{Jia}}, \citenamefont {{Dai}},
  \citenamefont {{Fang}}, \citenamefont {{Zhang}}, \citenamefont {{He}},
  \citenamefont {{Wang}}, \citenamefont {{Lu}}, \citenamefont {{Ma}},\ and\
  \citenamefont {{Xue}}}]{QKXu2013}%
  \BibitemOpen
  \bibfield  {author} {\bibinfo {author} {\bibfnamefont {C.-Z.}\ \bibnamefont
  {{Chang}}}, \bibinfo {author} {\bibfnamefont {J.}~\bibnamefont {{Zhang}}},
  \bibinfo {author} {\bibfnamefont {X.}~\bibnamefont {{Feng}}}, \bibinfo
  {author} {\bibfnamefont {J.}~\bibnamefont {{Shen}}}, \bibinfo {author}
  {\bibfnamefont {Z.}~\bibnamefont {{Zhang}}}, \bibinfo {author} {\bibfnamefont
  {M.}~\bibnamefont {{Guo}}}, \bibinfo {author} {\bibfnamefont
  {K.}~\bibnamefont {{Li}}}, \bibinfo {author} {\bibfnamefont {Y.}~\bibnamefont
  {{Ou}}}, \bibinfo {author} {\bibfnamefont {P.}~\bibnamefont {{Wei}}},
  \bibinfo {author} {\bibfnamefont {L.-L.}\ \bibnamefont {{Wang}}}, \bibinfo
  {author} {\bibfnamefont {Z.-Q.}\ \bibnamefont {{Ji}}}, \bibinfo {author}
  {\bibfnamefont {Y.}~\bibnamefont {{Feng}}}, \bibinfo {author} {\bibfnamefont
  {S.}~\bibnamefont {{Ji}}}, \bibinfo {author} {\bibfnamefont {X.}~\bibnamefont
  {{Chen}}}, \bibinfo {author} {\bibfnamefont {J.}~\bibnamefont {{Jia}}},
  \bibinfo {author} {\bibfnamefont {X.}~\bibnamefont {{Dai}}}, \bibinfo
  {author} {\bibfnamefont {Z.}~\bibnamefont {{Fang}}}, \bibinfo {author}
  {\bibfnamefont {S.-C.}\ \bibnamefont {{Zhang}}}, \bibinfo {author}
  {\bibfnamefont {K.}~\bibnamefont {{He}}}, \bibinfo {author} {\bibfnamefont
  {Y.}~\bibnamefont {{Wang}}}, \bibinfo {author} {\bibfnamefont
  {L.}~\bibnamefont {{Lu}}}, \bibinfo {author} {\bibfnamefont {X.-C.}\
  \bibnamefont {{Ma}}},\ and\ \bibinfo {author} {\bibfnamefont {Q.-K.}\
  \bibnamefont {{Xue}}},\ }\bibfield  {title} {\bibinfo {title} {{Experimental
  Observation of the Quantum Anomalous Hall Effect in a Magnetic Topological
  Insulator}},\ }\href {https://doi.org/10.1126/science.1234414} {\bibfield
  {journal} {\bibinfo  {journal} {Science}\ }\textbf {\bibinfo {volume}
  {340}},\ \bibinfo {pages} {167} (\bibinfo {year} {2013})}\BibitemShut
  {NoStop}%
\bibitem [{\citenamefont {Chang}\ \emph {et~al.}(2015)\citenamefont {Chang},
  \citenamefont {Zhao}, \citenamefont {Kim}, \citenamefont {Zhang},
  \citenamefont {Assaf}, \citenamefont {Heiman}, \citenamefont {Zhang},
  \citenamefont {Liu}, \citenamefont {Chan},\ and\ \citenamefont
  {Moodera}}]{Chang2015}%
  \BibitemOpen
  \bibfield  {author} {\bibinfo {author} {\bibfnamefont {C.-Z.}\ \bibnamefont
  {Chang}}, \bibinfo {author} {\bibfnamefont {W.}~\bibnamefont {Zhao}},
  \bibinfo {author} {\bibfnamefont {D.~Y.}\ \bibnamefont {Kim}}, \bibinfo
  {author} {\bibfnamefont {H.}~\bibnamefont {Zhang}}, \bibinfo {author}
  {\bibfnamefont {B.~A.}\ \bibnamefont {Assaf}}, \bibinfo {author}
  {\bibfnamefont {D.}~\bibnamefont {Heiman}}, \bibinfo {author} {\bibfnamefont
  {S.-C.}\ \bibnamefont {Zhang}}, \bibinfo {author} {\bibfnamefont
  {C.}~\bibnamefont {Liu}}, \bibinfo {author} {\bibfnamefont {M.~H.~W.}\
  \bibnamefont {Chan}},\ and\ \bibinfo {author} {\bibfnamefont {J.~S.}\
  \bibnamefont {Moodera}},\ }\bibfield  {title} {\bibinfo {title}
  {{High-precision realization of robust quantum anomalous Hall state in a hard
  ferromagnetic topological insulator}},\ }\href
  {https://doi.org/10.1038/nmat4204} {\bibfield  {journal} {\bibinfo  {journal}
  {Nature Mater}\ }\textbf {\bibinfo {volume} {14}},\ \bibinfo {pages} {473}
  (\bibinfo {year} {2015})}\BibitemShut {NoStop}%
\bibitem [{\citenamefont {Ko\"onig}\ \emph {et~al.}(2007)\citenamefont
  {Ko\"onig}, \citenamefont {Wiedmann}, \citenamefont {Br\"une}, \citenamefont
  {Roth}, \citenamefont {Buhmann}, \citenamefont {Molenkamp}, \citenamefont
  {Qi},\ and\ \citenamefont {Zhang}}]{Konig2007}%
  \BibitemOpen
  \bibfield  {author} {\bibinfo {author} {\bibfnamefont {M.}~\bibnamefont
  {Ko\"onig}}, \bibinfo {author} {\bibfnamefont {S.}~\bibnamefont {Wiedmann}},
  \bibinfo {author} {\bibfnamefont {C.}~\bibnamefont {Br\"une}}, \bibinfo
  {author} {\bibfnamefont {A.}~\bibnamefont {Roth}}, \bibinfo {author}
  {\bibfnamefont {H.}~\bibnamefont {Buhmann}}, \bibinfo {author} {\bibfnamefont
  {L.~W.}\ \bibnamefont {Molenkamp}}, \bibinfo {author} {\bibfnamefont {X.-L.}\
  \bibnamefont {Qi}},\ and\ \bibinfo {author} {\bibfnamefont {S.-C.}\
  \bibnamefont {Zhang}},\ }\bibfield  {title} {\bibinfo {title} {{Quantum Spin
  Hall Insulator State in {HgTe} Quantum Wells}},\ }\href
  {https://doi.org/10.1126/science.1148047} {\bibfield  {journal} {\bibinfo
  {journal} {Science}\ }\textbf {\bibinfo {volume} {318}},\ \bibinfo {pages}
  {766} (\bibinfo {year} {2007})}\BibitemShut {NoStop}%
\bibitem [{\citenamefont {Knez}\ \emph {et~al.}(2011)\citenamefont {Knez},
  \citenamefont {Du},\ and\ \citenamefont {Sullivan}}]{Knez2011}%
  \BibitemOpen
  \bibfield  {author} {\bibinfo {author} {\bibfnamefont {I.}~\bibnamefont
  {Knez}}, \bibinfo {author} {\bibfnamefont {R.-R.}\ \bibnamefont {Du}},\ and\
  \bibinfo {author} {\bibfnamefont {G.}~\bibnamefont {Sullivan}},\ }\bibfield
  {title} {\bibinfo {title} {{Evidence for Helical Edge Modes in Inverted
  $\mathrm{InAs}/\mathrm{GaSb}$ Quantum Wells}},\ }\href
  {https://doi.org/10.1103/PhysRevLett.107.136603} {\bibfield  {journal}
  {\bibinfo  {journal} {Phys. Rev. Lett.}\ }\textbf {\bibinfo {volume} {107}},\
  \bibinfo {pages} {136603} (\bibinfo {year} {2011})}\BibitemShut {NoStop}%
\bibitem [{\citenamefont {{Krishana, K and Ong, N. P. and Li, Q. and Gu, G. D.
  and Koshizuka, N.}}(1997)}]{Krishana1997}%
  \BibitemOpen
  \bibfield  {author} {\bibinfo {author} {\bibnamefont {{Krishana, K and Ong,
  N. P. and Li, Q. and Gu, G. D. and Koshizuka, N.}}},\ }\bibfield  {title}
  {\bibinfo {title} {{Plateaus Observed in the Field Profile of Thermal
  Conductivity in the Superconductor Bi$_2$Sr$_2$CaCu$_2$O$_8$}},\ }\href
  {https://doi.org/10.1126/science.277.5322.83} {\bibfield  {journal} {\bibinfo
   {journal} {Science}\ }\textbf {\bibinfo {volume} {277}},\ \bibinfo {pages}
  {83} (\bibinfo {year} {1997})}\BibitemShut {NoStop}%
\bibitem [{\citenamefont {Movshovich}\ \emph {et~al.}(1998)\citenamefont
  {Movshovich}, \citenamefont {Hubbard}, \citenamefont {Salamon}, \citenamefont
  {Balatsky}, \citenamefont {Yoshizaki}, \citenamefont {Sarrao},\ and\
  \citenamefont {Jaime}}]{Movshovich1998}%
  \BibitemOpen
  \bibfield  {author} {\bibinfo {author} {\bibfnamefont {R.}~\bibnamefont
  {Movshovich}}, \bibinfo {author} {\bibfnamefont {M.~A.}\ \bibnamefont
  {Hubbard}}, \bibinfo {author} {\bibfnamefont {M.~B.}\ \bibnamefont
  {Salamon}}, \bibinfo {author} {\bibfnamefont {A.~V.}\ \bibnamefont
  {Balatsky}}, \bibinfo {author} {\bibfnamefont {R.}~\bibnamefont {Yoshizaki}},
  \bibinfo {author} {\bibfnamefont {J.~L.}\ \bibnamefont {Sarrao}},\ and\
  \bibinfo {author} {\bibfnamefont {M.}~\bibnamefont {Jaime}},\ }\bibfield
  {title} {\bibinfo {title} {{Low-Temperature Anomaly in Thermal Conductivity
  of
  ${\mathrm{Bi}}_{2}{\mathrm{Sr}}_{2}\mathrm{Ca}({\mathrm{Cu}}_{1\ensuremath{-}\mathit{x}}{\mathrm{Ni}}_{\mathit{x}}{)}_{2}{\mathrm{O}}_{8}$:
  Second Superconducting Phase?}},\ }\href
  {https://doi.org/10.1103/PhysRevLett.80.1968} {\bibfield  {journal} {\bibinfo
   {journal} {Phys. Rev. Lett.}\ }\textbf {\bibinfo {volume} {80}},\ \bibinfo
  {pages} {1968} (\bibinfo {year} {1998})}\BibitemShut {NoStop}%
\bibitem [{\citenamefont {{Laughlin, R. B.}}(1998)}]{Laughlin1998}%
  \BibitemOpen
  \bibfield  {author} {\bibinfo {author} {\bibnamefont {{Laughlin, R. B.}}},\
  }\bibfield  {title} {\bibinfo {title} {{Magnetic Induction of
  ${\mathit{d}}_{{\mathit{x}}^{2}\ensuremath{-}{\mathit{y}}^{2}}+{\mathit{id}}_{\mathit{xy}}$
  Order in High- ${T}_{c}$ Superconductors}},\ }\href
  {https://doi.org/10.1103/PhysRevLett.80.5188} {\bibfield  {journal} {\bibinfo
   {journal} {Phys. Rev. Lett.}\ }\textbf {\bibinfo {volume} {80}},\ \bibinfo
  {pages} {5188} (\bibinfo {year} {1998})}\BibitemShut {NoStop}%
\bibitem [{\citenamefont {{Zhao}}\ \emph {et~al.}()\citenamefont {{Zhao}},
  \citenamefont {{Poccia}}, \citenamefont {{Cui}}, \citenamefont {{Volkov}},
  \citenamefont {{Yoo}}, \citenamefont {{Engelke}}, \citenamefont {{Ronen}},
  \citenamefont {{Zhong}}, \citenamefont {{Gu}}, \citenamefont {{Plugge}},
  \citenamefont {{Tummuru}}, \citenamefont {{Franz}}, \citenamefont
  {{Pixley}},\ and\ \citenamefont {{Kim}}}]{Pkim2021}%
  \BibitemOpen
  \bibfield  {author} {\bibinfo {author} {\bibfnamefont {S.~Y.~F.}\
  \bibnamefont {{Zhao}}}, \bibinfo {author} {\bibfnamefont {N.}~\bibnamefont
  {{Poccia}}}, \bibinfo {author} {\bibfnamefont {X.}~\bibnamefont {{Cui}}},
  \bibinfo {author} {\bibfnamefont {P.~A.}\ \bibnamefont {{Volkov}}}, \bibinfo
  {author} {\bibfnamefont {H.}~\bibnamefont {{Yoo}}}, \bibinfo {author}
  {\bibfnamefont {R.}~\bibnamefont {{Engelke}}}, \bibinfo {author}
  {\bibfnamefont {Y.}~\bibnamefont {{Ronen}}}, \bibinfo {author} {\bibfnamefont
  {R.}~\bibnamefont {{Zhong}}}, \bibinfo {author} {\bibfnamefont
  {G.}~\bibnamefont {{Gu}}}, \bibinfo {author} {\bibfnamefont {S.}~\bibnamefont
  {{Plugge}}}, \bibinfo {author} {\bibfnamefont {T.}~\bibnamefont {{Tummuru}}},
  \bibinfo {author} {\bibfnamefont {M.}~\bibnamefont {{Franz}}}, \bibinfo
  {author} {\bibfnamefont {J.~H.}\ \bibnamefont {{Pixley}}},\ and\ \bibinfo
  {author} {\bibfnamefont {P.}~\bibnamefont {{Kim}}},\ }\href@noop {} {\bibinfo
  {title} {{Emergent Interfacial Superconductivity between Twisted Cuprate
  Superconductors}}},\ \Eprint {https://arxiv.org/abs/arXiv:2108.13455}
  {arXiv:2108.13455} \BibitemShut {NoStop}%
\bibitem [{\citenamefont {Fu}(2011)}]{Fu2011}%
  \BibitemOpen
  \bibfield  {author} {\bibinfo {author} {\bibfnamefont {L.}~\bibnamefont
  {Fu}},\ }\bibfield  {title} {\bibinfo {title} {{Topological Crystalline
  Insulators}},\ }\href {https://doi.org/10.1103/PhysRevLett.106.106802}
  {\bibfield  {journal} {\bibinfo  {journal} {Phys. Rev. Lett.}\ }\textbf
  {\bibinfo {volume} {106}},\ \bibinfo {pages} {106802} (\bibinfo {year}
  {2011})}\BibitemShut {NoStop}%
\bibitem [{\citenamefont {Slager}\ \emph {et~al.}(2012)\citenamefont {Slager},
  \citenamefont {Mesaros}, \citenamefont {Juri{\v{c}}i{\'{c}}},\ and\
  \citenamefont {Zaanen}}]{Slager2012}%
  \BibitemOpen
  \bibfield  {author} {\bibinfo {author} {\bibfnamefont {R.-J.}\ \bibnamefont
  {Slager}}, \bibinfo {author} {\bibfnamefont {A.}~\bibnamefont {Mesaros}},
  \bibinfo {author} {\bibfnamefont {V.}~\bibnamefont {Juri{\v{c}}i{\'{c}}}},\
  and\ \bibinfo {author} {\bibfnamefont {J.}~\bibnamefont {Zaanen}},\
  }\bibfield  {title} {\bibinfo {title} {{The space group classification of
  topological band-insulators}},\ }\href {https://doi.org/10.1038/nphys2513}
  {\bibfield  {journal} {\bibinfo  {journal} {Nat. Phys.}\ }\textbf {\bibinfo
  {volume} {9}},\ \bibinfo {pages} {98} (\bibinfo {year} {2012})}\BibitemShut
  {NoStop}%
\bibitem [{\citenamefont {Shiozaki}\ and\ \citenamefont
  {Sato}(2014)}]{Shiozaki2014}%
  \BibitemOpen
  \bibfield  {author} {\bibinfo {author} {\bibfnamefont {K.}~\bibnamefont
  {Shiozaki}}\ and\ \bibinfo {author} {\bibfnamefont {M.}~\bibnamefont
  {Sato}},\ }\bibfield  {title} {\bibinfo {title} {{Topology of crystalline
  insulators and superconductors}},\ }\href
  {https://doi.org/10.1103/PhysRevB.90.165114} {\bibfield  {journal} {\bibinfo
  {journal} {Phys. Rev. B}\ }\textbf {\bibinfo {volume} {90}},\ \bibinfo
  {pages} {165114} (\bibinfo {year} {2014})}\BibitemShut {NoStop}%
\bibitem [{\citenamefont {{Bradlyn}}\ \emph {et~al.}(2017)\citenamefont
  {{Bradlyn}}, \citenamefont {{Elcoro}}, \citenamefont {{Cano}}, \citenamefont
  {{Vergniory}}, \citenamefont {{Wang}}, \citenamefont {{Felser}},
  \citenamefont {{Aroyo}},\ and\ \citenamefont {{Bernevig}}}]{bernevig2017}%
  \BibitemOpen
  \bibfield  {author} {\bibinfo {author} {\bibfnamefont {B.}~\bibnamefont
  {{Bradlyn}}}, \bibinfo {author} {\bibfnamefont {L.}~\bibnamefont {{Elcoro}}},
  \bibinfo {author} {\bibfnamefont {J.}~\bibnamefont {{Cano}}}, \bibinfo
  {author} {\bibfnamefont {M.~G.}\ \bibnamefont {{Vergniory}}}, \bibinfo
  {author} {\bibfnamefont {Z.}~\bibnamefont {{Wang}}}, \bibinfo {author}
  {\bibfnamefont {C.}~\bibnamefont {{Felser}}}, \bibinfo {author}
  {\bibfnamefont {M.~I.}\ \bibnamefont {{Aroyo}}},\ and\ \bibinfo {author}
  {\bibfnamefont {B.~A.}\ \bibnamefont {{Bernevig}}},\ }\bibfield  {title}
  {\bibinfo {title} {{Topological quantum chemistry}},\ }\href
  {https://doi.org/10.1038/nature23268} {\bibfield  {journal} {\bibinfo
  {journal} {\nat}\ }\textbf {\bibinfo {volume} {547}},\ \bibinfo {pages} {298}
  (\bibinfo {year} {2017})}\BibitemShut {NoStop}%
\bibitem [{\citenamefont {{Po}}\ \emph {et~al.}(2017)\citenamefont {{Po}},
  \citenamefont {{Vishwanath}},\ and\ \citenamefont
  {{Watanabe}}}]{Vishwanath2017NatComm}%
  \BibitemOpen
  \bibfield  {author} {\bibinfo {author} {\bibfnamefont {H.~C.}\ \bibnamefont
  {{Po}}}, \bibinfo {author} {\bibfnamefont {A.}~\bibnamefont {{Vishwanath}}},\
  and\ \bibinfo {author} {\bibfnamefont {H.}~\bibnamefont {{Watanabe}}},\
  }\bibfield  {title} {\bibinfo {title} {{Complete theory of symmetry-based
  indicators of band topology}},\ }\href
  {https://doi.org/10.1038/s41467-017-00133-2} {\bibfield  {journal} {\bibinfo
  {journal} {Nat. Commun.}\ }\textbf {\bibinfo {volume} {8}},\ \bibinfo {eid}
  {50} (\bibinfo {year} {2017})}\BibitemShut {NoStop}%
\bibitem [{\citenamefont {Po}\ \emph {et~al.}(2018)\citenamefont {Po},
  \citenamefont {Watanabe},\ and\ \citenamefont {Vishwanath}}]{ashvin2018}%
  \BibitemOpen
  \bibfield  {author} {\bibinfo {author} {\bibfnamefont {H.~C.}\ \bibnamefont
  {Po}}, \bibinfo {author} {\bibfnamefont {H.}~\bibnamefont {Watanabe}},\ and\
  \bibinfo {author} {\bibfnamefont {A.}~\bibnamefont {Vishwanath}},\ }\bibfield
   {title} {\bibinfo {title} {{Fragile Topology and Wannier Obstructions}},\
  }\href {https://doi.org/10.1103/PhysRevLett.121.126402} {\bibfield  {journal}
  {\bibinfo  {journal} {Phys. Rev. Lett.}\ }\textbf {\bibinfo {volume} {121}},\
  \bibinfo {pages} {126402} (\bibinfo {year} {2018})}\BibitemShut {NoStop}%
\bibitem [{\citenamefont {Zhang}\ \emph {et~al.}(2019)\citenamefont {Zhang},
  \citenamefont {Jiang}, \citenamefont {Song}, \citenamefont {Huang},
  \citenamefont {He}, \citenamefont {Fang}, \citenamefont {Weng},\ and\
  \citenamefont {Fang}}]{Zhang2019}%
  \BibitemOpen
  \bibfield  {author} {\bibinfo {author} {\bibfnamefont {T.}~\bibnamefont
  {Zhang}}, \bibinfo {author} {\bibfnamefont {Y.}~\bibnamefont {Jiang}},
  \bibinfo {author} {\bibfnamefont {Z.}~\bibnamefont {Song}}, \bibinfo {author}
  {\bibfnamefont {H.}~\bibnamefont {Huang}}, \bibinfo {author} {\bibfnamefont
  {Y.}~\bibnamefont {He}}, \bibinfo {author} {\bibfnamefont {Z.}~\bibnamefont
  {Fang}}, \bibinfo {author} {\bibfnamefont {H.}~\bibnamefont {Weng}},\ and\
  \bibinfo {author} {\bibfnamefont {C.}~\bibnamefont {Fang}},\ }\bibfield
  {title} {\bibinfo {title} {{Catalogue of topological electronic materials}},\
  }\href {https://doi.org/10.1038/s41586-019-0944-6} {\bibfield  {journal}
  {\bibinfo  {journal} {\nat}\ }\textbf {\bibinfo {volume} {566}},\ \bibinfo
  {pages} {475} (\bibinfo {year} {2019})}\BibitemShut {NoStop}%
\bibitem [{\citenamefont {Vergniory}\ \emph {et~al.}(2019)\citenamefont
  {Vergniory}, \citenamefont {Elcoro}, \citenamefont {Felser}, \citenamefont
  {Regnault}, \citenamefont {Bernevig},\ and\ \citenamefont
  {Wang}}]{Vergniory2019}%
  \BibitemOpen
  \bibfield  {author} {\bibinfo {author} {\bibfnamefont {M.~G.}\ \bibnamefont
  {Vergniory}}, \bibinfo {author} {\bibfnamefont {L.}~\bibnamefont {Elcoro}},
  \bibinfo {author} {\bibfnamefont {C.}~\bibnamefont {Felser}}, \bibinfo
  {author} {\bibfnamefont {N.}~\bibnamefont {Regnault}}, \bibinfo {author}
  {\bibfnamefont {B.~A.}\ \bibnamefont {Bernevig}},\ and\ \bibinfo {author}
  {\bibfnamefont {Z.}~\bibnamefont {Wang}},\ }\bibfield  {title} {\bibinfo
  {title} {{A complete catalogue of high-quality topological materials}},\
  }\href {https://doi.org/10.1038/s41586-019-0954-4} {\bibfield  {journal}
  {\bibinfo  {journal} {\nat}\ }\textbf {\bibinfo {volume} {566}},\ \bibinfo
  {pages} {480} (\bibinfo {year} {2019})}\BibitemShut {NoStop}%
\bibitem [{\citenamefont {Tang}\ \emph {et~al.}(2019)\citenamefont {Tang},
  \citenamefont {Po}, \citenamefont {Vishwanath},\ and\ \citenamefont
  {Wan}}]{Tang2019}%
  \BibitemOpen
  \bibfield  {author} {\bibinfo {author} {\bibfnamefont {F.}~\bibnamefont
  {Tang}}, \bibinfo {author} {\bibfnamefont {H.~C.}\ \bibnamefont {Po}},
  \bibinfo {author} {\bibfnamefont {A.}~\bibnamefont {Vishwanath}},\ and\
  \bibinfo {author} {\bibfnamefont {X.}~\bibnamefont {Wan}},\ }\bibfield
  {title} {\bibinfo {title} {{Comprehensive search for topological materials
  using symmetry indicators}},\ }\href
  {https://doi.org/10.1038/s41586-019-0937-5} {\bibfield  {journal} {\bibinfo
  {journal} {\nat}\ }\textbf {\bibinfo {volume} {566}},\ \bibinfo {pages} {486}
  (\bibinfo {year} {2019})}\BibitemShut {NoStop}%
\end{thebibliography}%

\end{document}